\documentclass[10pt,twocolumn,showkeys,amssymb,amsmath,aps,superscriptaddress]{revtex4-2}
\usepackage{graphicx}
\usepackage{xcolor}
\usepackage[colorlinks={true}]{hyperref}
\hypersetup{colorlinks=true,linkcolor=red,citecolor=blue,urlcolor=blue}
\newcommand{\orange}[1]{\textcolor{orange}{#1}}
\newcommand{\green}[1]{\textcolor{green}{#1}}

\begin{document}

\preprint{APS/123-QED}
\title{Fingerprints of cluster-based Haldane and bound-magnon states \\ in a spin-1 Heisenberg diamond chain}
\author{Azam Zoshki}
\affiliation{Department of Theoretical Physics and Astrophysics, Faculty of Science, P. J. \v{S}af{\'a}rik University, Park Angelinum 9, 040 01 Ko\v{s}ice, Slovak Republic}
\author{Hamid Arian Zad}
\affiliation{Department of Theoretical Physics and Astrophysics, Faculty of Science, P. J. \v{S}af{\'a}rik University, Park Angelinum 9, 040 01 Ko\v{s}ice, Slovak Republic}
\author{Katar\'ina Karl'ov\'a}
\affiliation{Department of Theoretical Physics and Astrophysics, Faculty of Science, P. J. \v{S}af{\'a}rik University, Park Angelinum 9, 040 01 Ko\v{s}ice, Slovak Republic}
\author{Jozef Stre\v{c}ka}
\email{Corresponding author: jozef.strecka@upjs.sk}
\affiliation{Department of Theoretical Physics and Astrophysics, Faculty of Science, P. J. \v{S}af{\'a}rik University, Park Angelinum 9, 040 01 Ko\v{s}ice, Slovak Republic}

\date{\today}

\begin{abstract}
We investigate magnetic and thermodynamic properties of a spin-1 Heisenberg diamond chain in a magnetic field using a combination of analytical and numerical methods including the variational approach, exact diagonalization, density-matrix renormalization group, localized-magnon theory, and quantum Monte Carlo simulations. In the unfrustrated regime, the model exhibits a quantum ferrimagnetic phase that captures key magnetic features of the nickel-based polymeric compound [Ni$_3$(OH)$_2$(C$_4$H$_2$O$_4$)(H$_2$O)$_4$] $\cdot$ 2H$_2$O such as a flat minimum in the temperature dependence of the susceptibility times temperature product and an intermediate one-third magnetization plateau. In the frustrated regime, we uncover a rich variety of unconventional quantum phases including uniform and cluster-based Haldane states, fragmented monomer–dimer phase, and bound-magnon crystals. Analysis of the adiabatic temperature change and magnetic Gr\"uneisen parameter reveals an enhanced magnetocaloric effect near field-induced transitions between these exotic quantum phases. Additionally, we demonstrate that the frustrated spin-1 diamond chain can operate as an efficient working medium of a quantum Stirling engine, which approaches near-optimal efficiency when driven into these unconventional quantum states. 
\end{abstract}

\maketitle

\section{Introduction} 
\label{sec:Introduction}

Frustrated quantum magnets represent a fascinating class of materials  in which competing interactions suppress conventional magnetic order and give rise to a broad spectrum of unconventional quantum phenomena \cite{lacr11}. They provide a versatile platform for realizing exotic quantum  ground states including the topologically nontrivial Haldane state \cite{hald83a,hald83b,affl87}, various quantum spin liquids \cite{bale10,sava17,gege20,chou25,bouc25}, bound-magnon crystals \cite{schu02,zhit05,derz06,derz15,okum19}, and fragmented \cite{shas81,bose89,bose90,bose92} phases. These unconventional states often leave striking fingerprints such as fractional plateaus in low-temperature magnetization curves \cite{momo00,miya03,hone04,meis07}. Beyond their fundamental significance, frustrated quantum spin systems are also technologically appealing as they can exhibit an enhanced magnetocaloric effect and thus serve as efficient refrigerant materials for magnetic cooling at cryogenic temperatures \cite{zhit03,zhit04,gsch05}. Recently, quantum spin systems have also emerged as promising working media for quantum heat engines capable of reaching efficiencies close to the ideal Carnot limit when operating in quantum Stirling \cite{purk22,cruz23,cast25} and Otto \cite{rabb23,alsu24,roja25} cycles. These features place frustrated quantum magnets at the forefront of quantum materials research bridging fundamental concepts with emergent quantum technologies.

Among the broad class of frustrated quantum spin structures, the spin-1/2 Heisenberg diamond chain has attracted considerable attention primarily due to its celebrated realization in the natural mineral azurite Cu$_3$(CO$_3$)$_2$(OH)$_2$ displaying several remarkable magnetic features \cite{kiku03,kiku04,kiku05a,kiku05b}. Careful theoretical modeling of azurite’s magnetic response has elucidated the  microscopic nature of the intermediate one-third plateau observed in low-temperature magnetization curves as well as the characteristic double-peak profiles in temperature dependence of both magnetic susceptibility and specific heat \cite{jesc11,hone11}. Notably, the azurite is not the sole experimental representative of the spin-1/2 Heisenberg diamond chain. The intriguing quantum spin-liquid ground states have been recently reported in a series of copper-based coordination polymers A$_3$Cu$_3$AlO$_2$(SO$_4$)$_4$ (A = K, Rb, and Cs) \cite{fuji15,mori17,fuji17}, [Cu$_3$(OH)$_2$(CH$_3$COO)$_2$(H$_2$O)$_4$](RSO$_3$)$_2$ (RSO$_3$ = organic sulfonate anions) \cite{fuji16}, and Cu$_3$ (CH$_3$COO)$_4$(OH)$_2$ $\cdot$ 5H$_2$O \cite{cuim19}, which afford distorted variants of the spin-1/2 Heisenberg diamond chain. Besides the copper-based family, the structural motif of the diamond chain with higher spin magnitude has been identified in the nickel(II)-based compound [Ni$_3$(OH)$_2$(C$_4$H$_2$O$_4$)(H$_2$O)$_4$] $\cdot$ 2H$_2$O hosting a spin-1 diamond chain \cite{guil02}, the cobalt(II)-based compound Co$_3$(OH)$_2$(C$_4$O$_4$)$_2$ $\cdot$ 3H$_2$O providing a spin-3/2 diamond chain \cite{mole11}, or even mixed-valent iron(II, III)-based compound [(CH$_3$)$_2$NH$_2$]$_3$[Fe$_3$O(OH)(SO$_4$)$_4$] affording a mixed spin-(2, 5/2) diamond chain \cite{soro20}.

Compared with their spin-1/2 counterparts, the Heisenberg diamond chains with higher spin magnitudes remain far less explored. A comprehensive ground-state analysis of the spin-1 Heisenberg diamond chain in zero magnetic field revealed a remarkably rich spectrum of unconventional quantum ground states stabilized by geometric spin frustration \cite{hida17}.  Subsequent studies further extended this analysis to include lattice distortions \cite{hida19} and bond alternation effects \cite{hida20}. The undistorted version of the spin-1 Heisenberg diamond chain without bond alternation hosts two distinct ferrimagnetic ground states in the frustration-free or weakly frustrated regimes before it enters into the topologically nontrivial Haldane phase in a moderately frustrated regime \cite{hida17}. In contrast, the highly frustrated parameter regime is dominated by a fragmented monomer-dimer phase. However, there is no direct transition between the Haldane phase and monomer-dimer phases, which are separated by a sequence of quantum phase transitions involving three distinct fragmented cluster-based Haldane phases. So far, the cluster-based Haldane phase has been experimentally identified only in the natural mineral fedotovite K$_2$Cu$_3$O(SO$_4$)$_3$ \cite{fuji18}. This exciting experimental finding triggered considerable attention towards searching other paradigmatic examples of quantum Heisenberg spin chains with the cluster-based Haldane ground states such as the octahedral chain \cite{karl19,karl22}, spin-cluster chains \cite{sugi20}, and triangular spin tubes \cite{sugi23}. 

\begin{figure}
\hspace{-0.4cm}
\includegraphics[width=0.45\textwidth]{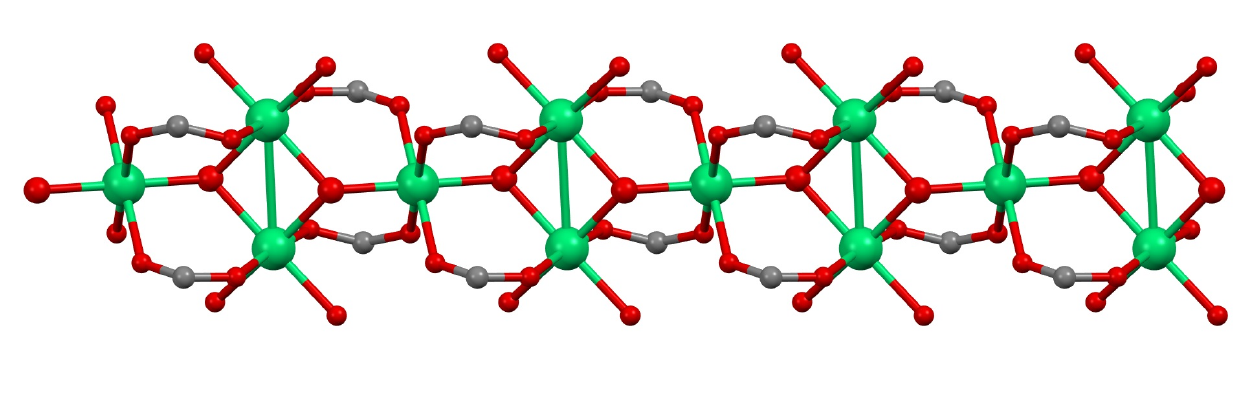}
\vspace{-0.8cm}
\caption{A part of the crystal structure of one-dimensional coordination polymer [Ni$_3$(OH)$_2$(C$_4$H$_2$O$_4$)(H$_2$O)$_4$] $\cdot$ 2H$_2$O visualized according to crystallographic data reported in Ref. \cite{guil02}. A color scheme for atom labeling: green balls - nickel, red balls - oxygen, gray balls - carbon.}
\label{fig:dcexp}
\end{figure}

The external magnetic field represents a key tuning parameter capable of controlling the stability of cluster-based Haldane phases. The primary goal of this work is to elucidate how the magnetic field influences the emergence and stability of cluster-based Haldane states in the spin-1 Heisenberg diamond chain, which undergoes fragmentation due to a periodically repeating pattern of dimer-singlet states formed on selected vertical bonds. From an experimental perspective, it regrettably turns out that the nickel-based polymeric compound [Ni$_3$(OH)$_2$(C$_4$H$_2$O$_4$)(H$_2$O)$_4$] $\cdot$ 2H$_2$O \cite{guil02} depicted in Fig. \ref{fig:dcexp} cannot host the cluster-based Haldane state as this practical realization of the spin-1 Heisenberg diamond chain falls into the unfrustrated regime. 

The fragmentation of the spin-1 Heisenberg diamond chain in the frustrated regime naturally suggests the feasibility of an effective lattice-gas formulation within the framework of the extended localized-magnon theory, which captures lowest-energy states of fragmented spin clusters as well as bound magnons allowing a unified and consistent description of low-temperature magnetic and thermodynamic properties. Unlike most frustrated quantum Heisenberg antiferromagnets, where the applicability of the localized-magnon theory is conventionally limited to high magnetic fields near saturation \cite{schu02,zhit05,derz06,derz15}, the fragmented nature of cluster-based Haldane states together with the bound character of magnons allows determination of low-temperature magnetic and thermodynamic properties down to zero field quite similarly as demonstrated for the spin-1/2 Heisenberg diamond and octahedral chains \cite{stre22}.   

The remainder of this paper is organized as follows. In Sec. \ref{HDC} we introduce the spin-1 Heisenberg diamond chain. The employed analytical and numerical techniques including the variational method, localized-magnon approach, density-matrix renormalization group, and exact diagonalization are outlined in Sec. \ref{MM}. The resulting ground-state phase diagram, which provides the backbone for the subsequent analysis, is established at the end of this section. Sec. \ref{FTM} addresses finite-temperature properties: we first analyze the magnetic susceptibility from exact diagonalization (ED) and quantum Monte Carlo (QMC) data, then confront theoretical predictions with experimental measurements for the magnetization and magnetic susceptibility of the nickel-based coordination polymer [Ni$_3$(OH)$_2$(C$_4$H$_2$O$_4$)(H$_2$O)$_4$] $\cdot$ 2H$_2$O and finally develop an effective lattice-gas description for the highly frustrated regime. In Sec. \ref{QSHE}, we demonstrate how the frustrated spin-1 Heisenberg diamond chain can serve as the working medium of a quantum Stirling heat engine. The main findings are summarized in Sec. \ref{CC}, where we also discuss their broader physical implications.

\section{Model}
\label{HDC}

\begin{figure}
\centering
\includegraphics[width=0.45\textwidth]{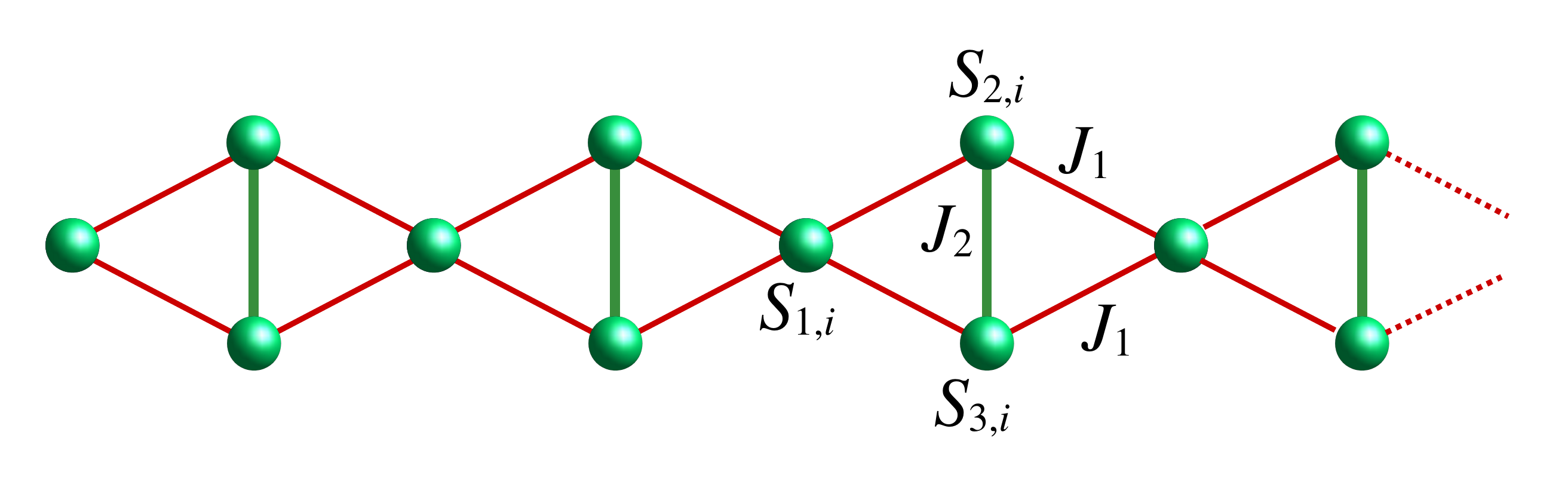}
\vspace{-0.75cm}
\caption{A schematic illustration of the spin-1 Heisenberg diamond chain with the intra-dimer interaction $J_2$ (green lines) along its vertical bonds and the monomer-dimer interaction $J_1$ (red lines) along sides of diamond plaquettes. Three spins from $i$-th unit cell are labeled.}
\label{fig:diamond_chain_model}
\end{figure}

Let us consider a symmetric spin-1 Heisenberg diamond chain with two distinct exchange interactions $J_1$ and $J_2$ as schematically illustrated in Fig. \ref{fig:diamond_chain_model}.
The Hamiltonian of the spin-1 Heisenberg diamond chain in the presence of an external magnetic field takes the form:
\begin{eqnarray}\label{Eq:hamiltonian}
\hat{\cal H}
&=& {J}_{1} \sum\limits_{i=1}^N \left(\hat{\boldsymbol S}_{1,i}+\hat{{\boldsymbol S}}_{1,i+1}\right) 
\cdot \left(\hat{{\boldsymbol S}}_{2,i}+\hat{{\boldsymbol S}}_{3,i}\right)
\nonumber \\
&+& {J}_{2}\sum\limits_{i=1}^N\hat{{\boldsymbol S}}_{2,i}\cdot\hat{{\boldsymbol S}}_{3,i}
-h\sum\limits_{i=1}^N\sum\limits_{j=1}^3\hat{S}_{j,i}^z,
\end{eqnarray}
where $\hat{\boldsymbol S}_{j,i}\equiv(\hat{S}_{j,i}^x,\hat{S}_{j,i}^y,\hat{S}_{j,i}^z)$ is the spin-1 operator associated with site $j=1,2,3$ of the $i$-th unit cell and $N$ denotes the total number of unit cells. The coupling constants $J_1$ and $J_2$ correspond to the monomer-dimer and intra-dimer interactions, whereby the last term $h=g\mu_{\mathrm{B}}H$ represents the standard Zeeman's term including the external magnetic field $H$, the Landé g-factor $g$, and the Bohr magneton $\mu_{\mathrm{B}}$. Since the model defined by the Hamiltonian (\ref{Eq:hamiltonian}) is not exactly solvable, we therefore employ a combination of complementary analytical and numerical techniques to explore its magnetic and thermodynamic properties. To minimize finite-size effects, periodic boundary conditions are imposed $S_{1,N+1} \equiv S_{1,1}$. 

\section{Methods}
\label{MM}

\subsection{Variational method}

\begin{center}
\begin{figure}
\includegraphics[width=0.45\textwidth]{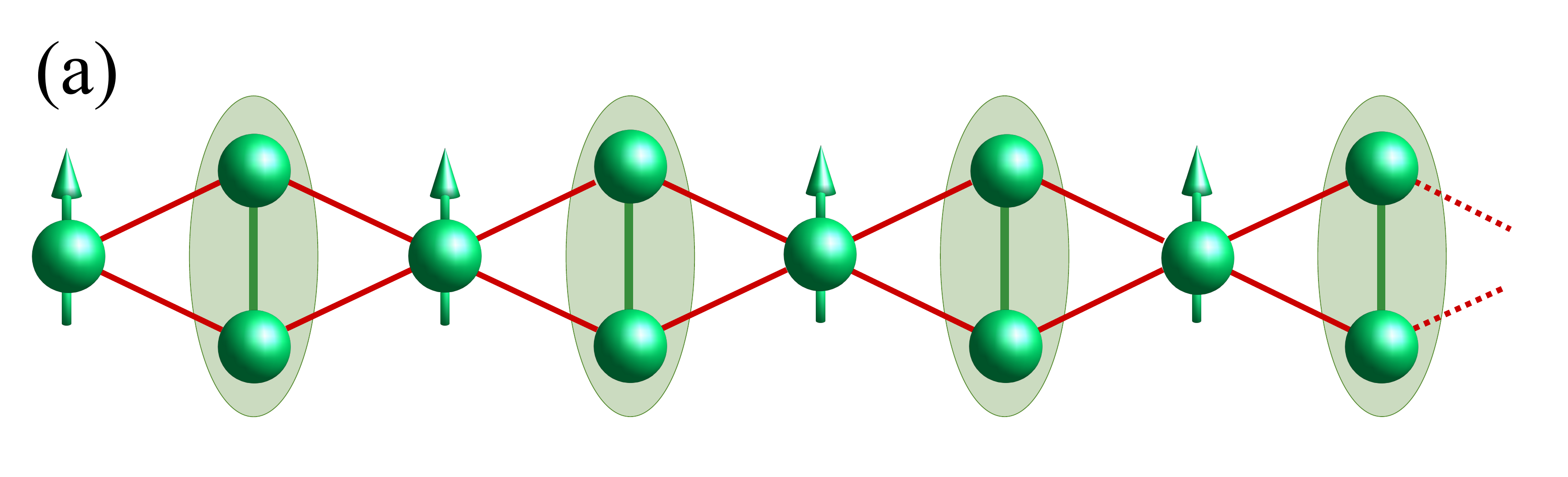}
\includegraphics[width=0.45\textwidth]{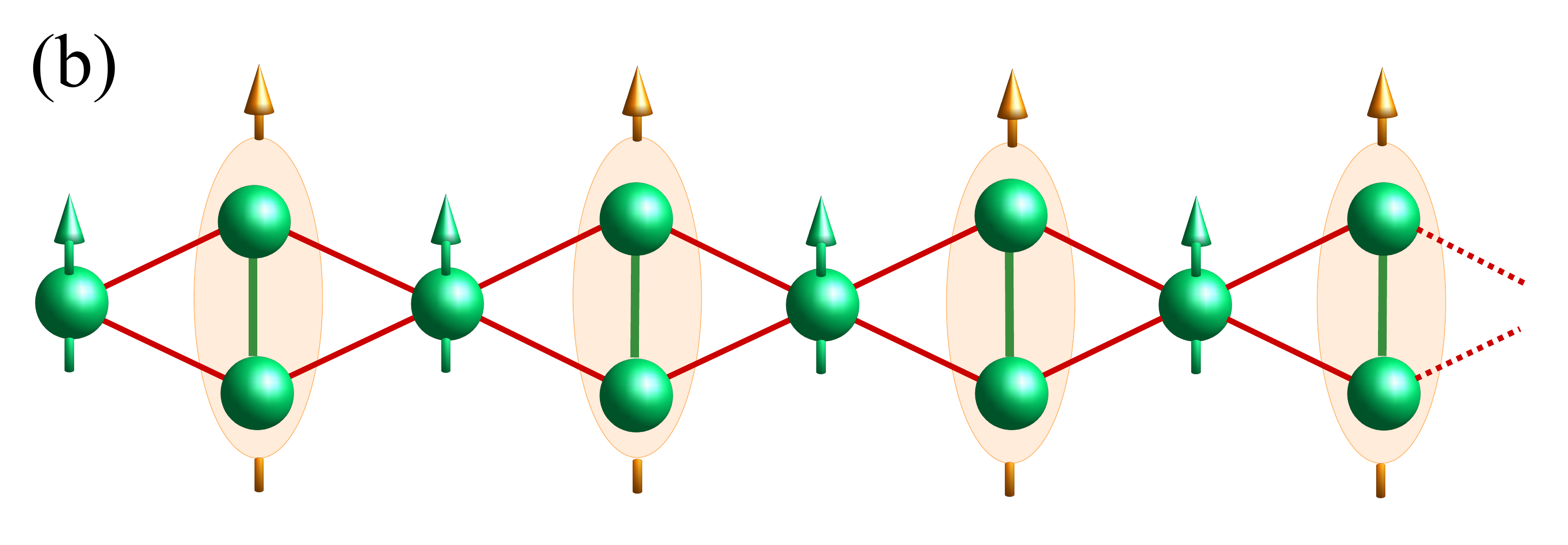} 
\includegraphics[width=0.45\textwidth]{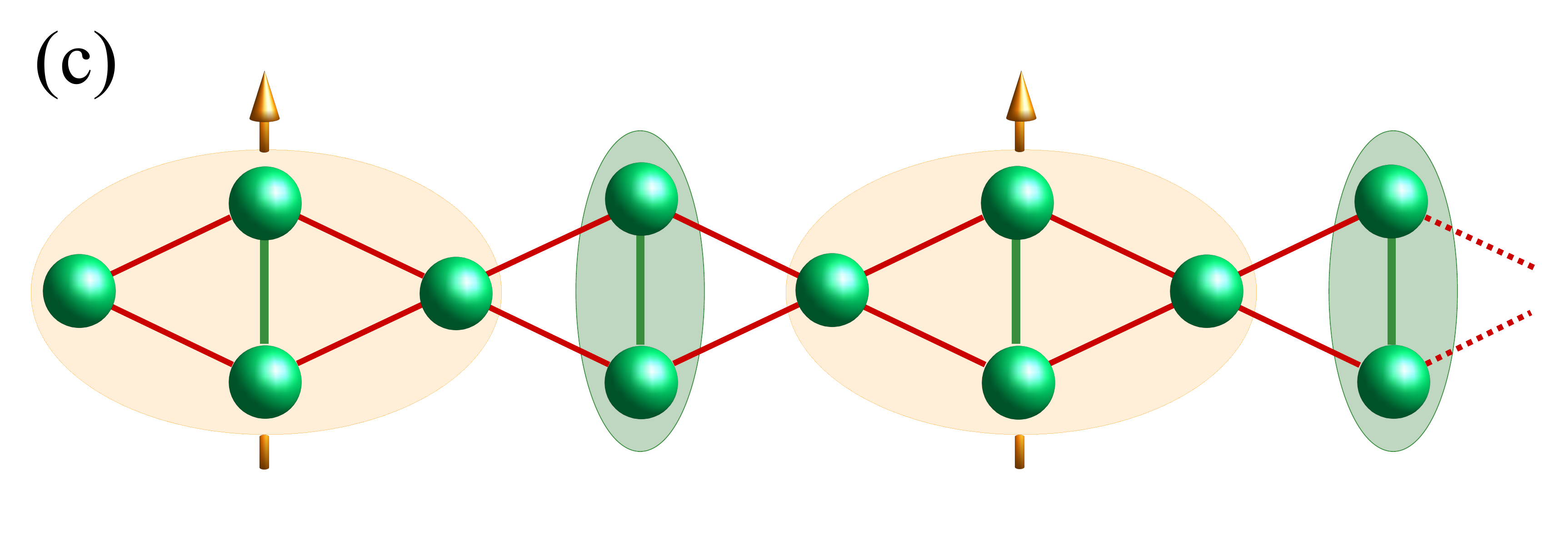}
\includegraphics[width=0.45\textwidth]{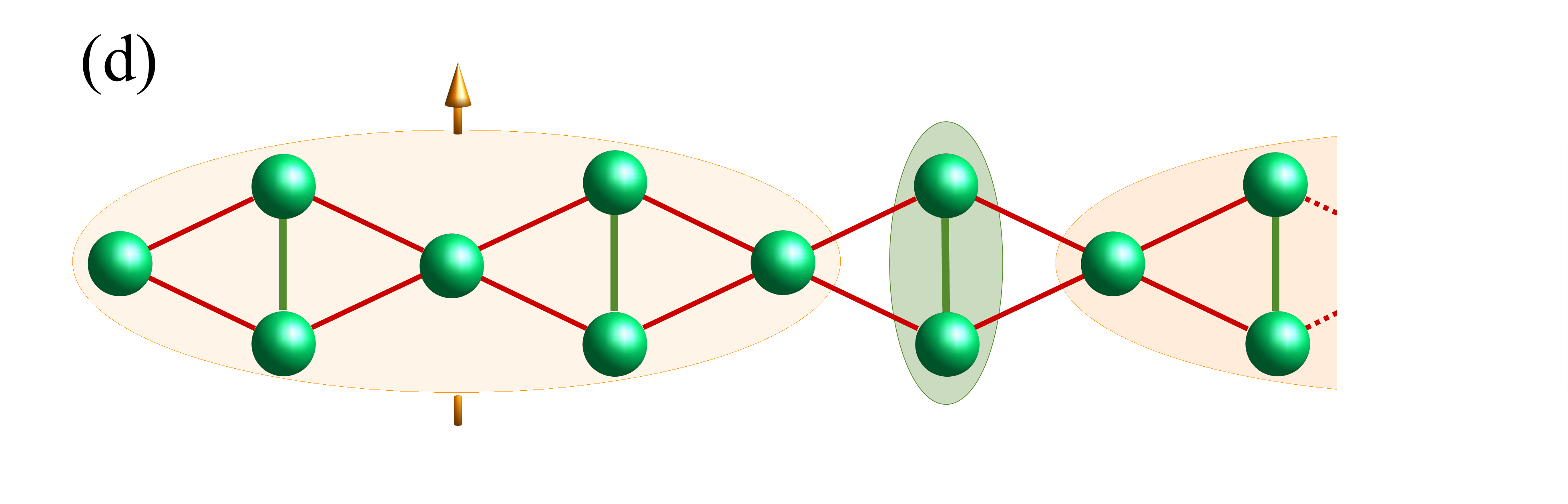}
\includegraphics[width=0.45\textwidth]{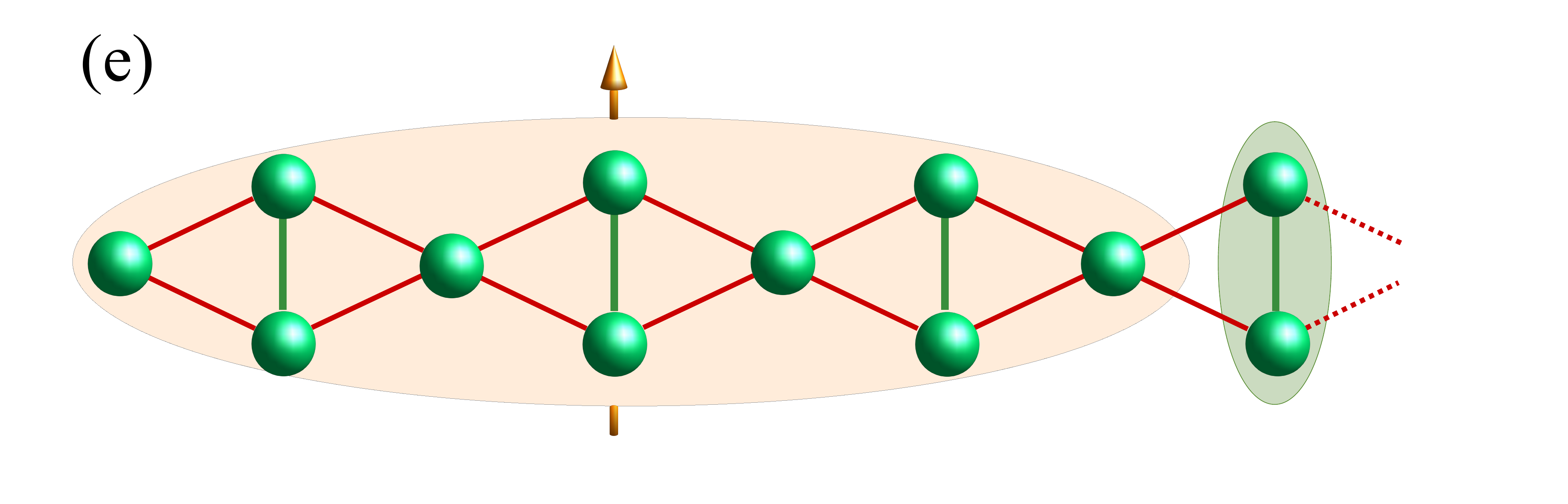}
\includegraphics[width=0.45\textwidth]{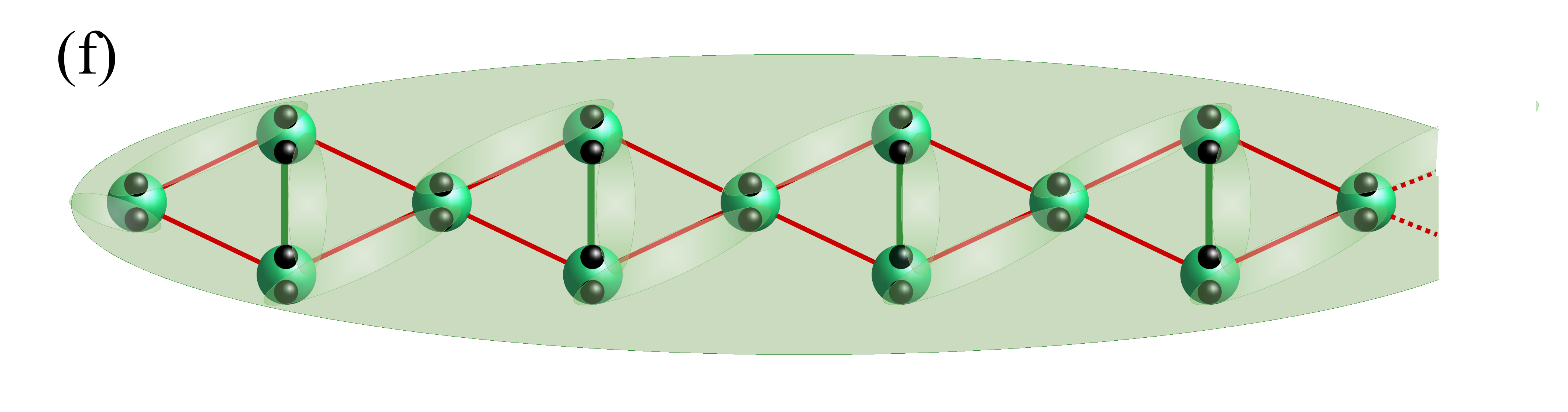}
\vspace{-0.5cm}
\caption{A schematic illustration of exact ground states of the spin-1 Heisenberg diamond chain: (a) the monomer-dimer (MD) phase; (b) the bound magnon crystal (BMC) phase; (c) the tetramer-dimer (TD) phase; (d) the heptamer-dimer (HD) phase; (e) the decamer-dimer (D-D) phase; (f) the Haldane phase. Green arrows correspond to the polarized monomer spins $S_{1,i}^z=+1$ [panels (a) and (b)], green ovals correspond to a dimer-singlet state [panels (a), (c), (d), and (e)], and orange ovals accompanied with orange arrows represent a triplet state of a dimer [panel (b)], a tetramer [panel (c)], a heptamer [panel (d)], or a decamer [panel (e)].}
\label{fig:SpinConfig}
\end{figure}
\end{center}

The variational principle due to Shastry and Sutherland provides a powerful route to determine exact ground states of frustrated quantum Heisenberg models within highly frustrated parameter regimes \cite{shas81}. In this framework, a rigorous lower bound of the ground-state energy is obtained as the sum of the lowest eigenenergies of suitably chosen subsystems into which the overall system is decomposed \cite{shas81,bose89,bose90,bose92}. The total Hamiltonian (\ref{Eq:hamiltonian}) of the spin-1 Heisenberg diamond chain can be eventually decomposed into the Hamiltonians of the triangular unit cells $\hat{\cal H} = \sum_{i=1}^N \sum_{\delta = 0, 1} \hat{\cal H}_{i,\delta} $ with the cell Hamiltonian defined as:
\begin{eqnarray}\label{H_eff_block_j}
\hat{\cal H}_{i,\delta} &=& J_{1} \hat{{\boldsymbol S}}_{1,i+\delta} \cdot (\hat{{\boldsymbol S}}_{2,i} + \hat{{\boldsymbol S}}_{3,i})  + \frac{J_{2}}{2} \hat{{\boldsymbol S}}_{2,i} \cdot \hat{{\boldsymbol S}}_{3,i} \nonumber\\ 
	&& -\frac{h}{2}(\hat{S}_{1,i+\delta}^z + \hat{S}_{2,i}^z + \hat{S}_{3,i}^z).
\end{eqnarray}
A most convenient diagonalization of the cell Hamiltonian (\ref{H_eff_block_j}) can be achieved by adopting the Kambe coupling scheme \cite{kamb50,sinn70}, in which the spin operators are rearranged into composite spins. Specifically, we introduce composite spin operator assigned to the total spin of $i$-th dimer \(\hat{{\boldsymbol S}}_{23, i} = \hat{{\boldsymbol S}}_{2,i} + \hat{{\boldsymbol S}}_{3,i}\), the total spin of triangular unit cell \(\hat{{\boldsymbol S}}_{t,i+\delta} = \hat{{\boldsymbol S}}_{1,i+\delta} + \hat{{\boldsymbol S}}_{2,i} + \hat{{\boldsymbol S}}_{3,i}\) and its $z$-component 
\(\hat{S}_{t,i+\delta}^z\) in terms of which the cell Hamiltonian takes the compact form:
\begin{equation}\label{Eq:Ham_DS1_KP}
\hat{\cal H}_{i,\delta} = -J_{1} - J_{2} + \frac{J_{1}}{2}\hat{\boldsymbol{S}}_{t,i+\delta}^2 
+ \left(\frac{J_{2}}{2} - J_{1}\right) \frac{\hat{\boldsymbol{S}}_{23,i}^{2}}{2} 
- \frac{h}{2}{\hat{S}_{t,i+\delta}^z}.
\end{equation}
The composite spin operators $\hat{{\boldsymbol S}}_{23,i}^2$, $\hat{{\boldsymbol S}}_{t,i}^2$, and $\hat{S}_{t,i}^z$ apparently commute with the cell Hamiltonian $\hat{\cal H}_{i,\delta}$ given by Eq. (\ref{Eq:Ham_DS1_KP}), which enables one to express respective energy eigenvalues directly in terms of the associated quantum spin numbers:
\begin{eqnarray}\label{Eq:Energies_DS1_KP}
\varepsilon_{i,\delta} &=& -{J}_{1} - {J}_{2} 
+ \frac{J_{1}}{2} {S}_{t,i+\delta}({S}_{t,i+\delta}+1) \nonumber \\ 
&+& \left(\frac{J_{2}}{4} - \frac{J_{1}}{2}\right){S}_{23,i}({S}_{23,i}+1) - \frac{h}{2}{S}_{t,i+\delta}^z.
\end{eqnarray}
According to the variational principle \cite{shas81,bose89,bose90,bose92}, the lower bound of the exact ground-state energy $E_0$ is determined by the sum of the lowest eigenenergies $\varepsilon_{i,\delta}^{(0)}$ of the triangular unit cells:
\begin{equation}
E_0 \geq \sum_{i=1}^{N} \sum_{\delta=0,1} \varepsilon_{i,\delta}^{(0)}.
\end{equation}
Employing the standard composition rules for quantum spin numbers ${S}_{23,i}$ = 0, 1 or 2, ${S}_{t,i+\delta}=\vert{S}_{23,i} - 1\vert, \vert{S}_{23,i} - 1\vert + 1, \cdots, {S}_{23,i} + 1$, and ${S}_{t,i+\delta}^z = - S_{t,i+\delta}, - S_{t,i+\delta} + 1, \cdots, {S}_{t,i+\delta}$, the full energy spectrum of the cell Hamiltonian (\ref{H_eff_block_j}) follows directly from Eq. (\ref{Eq:Energies_DS1_KP}). Inspection of this energy spectrum reveals that the eigenstate with the energy $\varepsilon_{i,\delta}^{(0)} = -J_2 - h/2$, which is obtained for the particular choice of the composite quantum spin numbers ${S}_{23,i} = 0$ and ${S}_{t,i+\delta} = {S}_{t,i+\delta}^z = 1$, represents the lowest-energy eigenstate in the parameter region $J_2 > 2J_1$, $h>2J_1-J_2/2$, and $h<2J_1+J_2$. The specific value of the composite quantum spin number ${S}_{23,i} = 0$ signifies the formation of a singlet dimer on the vertical $J_2$-bond and hence, this local eigenstate can be trivially extended to the entire spin-1 diamond chain to realize the monomer-dimer (MD) phase illustrated in Fig. \ref{fig:SpinConfig}(a) and given by the exact eigenvector:
\begin{eqnarray}\label{Eq:M-D}
\vert\text{MD}\rangle &=&\prod_{i=1}^{N}
\vert\!+\!1\rangle_{1,i}\dfrac{1}{\sqrt{3}}\big(\vert\!+\!1\rangle_{2,i}\vert\!-\!1\rangle_{3,i} 
+ \vert\!-\!1\rangle_{2,i} \vert\!+\!1\rangle_{3,i}\nonumber \\ 
&&\qquad\qquad\qquad -\vert 0\rangle_{2,i}\vert 0\rangle_{3,i}\big).
\end{eqnarray}
It can be actually proved that the MD phase (\ref{Eq:M-D}) is an exact eigenstate of the full spin-1 Heisenberg diamond chain with the overall energy $E_{\rm MD} = 2N \varepsilon_{i,\delta}^{(0)} = -2NJ_2 - Nh$, which exactly coincides with the lower bound of the variational energy in the parameter domain $J_2 > 2J_1$, $h>2J_1-J_2/2$, and $h<2J_1+J_2$. This establishes the  MD phase (\ref{Eq:M-D}) as an exact ground state of the spin-1 Heisenberg diamond chain at least in this parameter regime.

\subsection{Bound magnon crystal}

The spin-1 Heisenberg diamond chain naturally displays at sufficiently high magnetic fields the fully polarized ferromagnetic state:
\begin{eqnarray}
\label{Eq:FM-State}
\vert\text{FM}\rangle=\prod_{l=1}^{N}\vert +1\rangle_{1,l}\vert +1\rangle_{2,l}\vert +1\rangle_{3,l},
\end{eqnarray}
which represents another exact eigenstate of the Hamiltonian (\ref{Eq:hamiltonian}) with the energy $E_\text{FM} = 4NJ_1 + NJ_2 - 3Nh$.  In the highly frustrated regime, destructive quantum interference may stabilize the crystallization of bound magnons on the vertical dimers ($J_2$-bonds) below the saturation field \cite{derz06,derz15}. To explore a possible bound-magnon crystal (BMC) ground state, we solve the time-independent Schr\"odinger equation $\hat{\cal H}\vert\Psi\rangle = E\vert\Psi\rangle$ in one-magnon sector, where exact one-magnon eigenstates $\vert\Psi\rangle$ are constructed as quantum superposition of basis states $\vert j,l\rangle = \frac{1}{\sqrt{2}}\hat{S}_{j,l}^-\vert\text{FM}\rangle$ ($j=1-3$ and $l=1-N$) with a single spin-flip deviation from the fully polarized ferromagnetic state $\vert\text{FM}\rangle$:
\begin{eqnarray}
\label{Eq:Psi}
\vert\Psi\rangle=\sum_{l=1}^{N} \sum_{j=1}^{3} c_{j, l} \vert{j,l}\rangle.
\end{eqnarray}
Applying the Hamiltonian (\ref{Eq:hamiltonian}) to the one-magnon basis states $\vert j,l\rangle$ yields the following outcomes: 
\begin{eqnarray}\label{Eq:HPsi}
&& \hat{\cal H}\vert{1,l}\rangle=(E_\text{FM}-4J_1+h)\vert{1,l}\rangle + J_1\big(\vert{2,l-1}\rangle \nonumber
\\
&& \qquad\qquad\quad +\vert{3,l-1}\rangle +\vert{2,l}\rangle+\vert{3,l}\rangle\big), \nonumber
\\
&& \hat{\cal H}\vert{2,l}\rangle=(E_\text{FM}-2J_1-J_2+h)\vert{2,l}\rangle + J_2\vert{3,l}\rangle \nonumber
\\
&& \qquad\qquad\quad +J_1\big(\vert{1,l}\rangle +\vert{1,l+1}\rangle\big), \nonumber
\\
&& \hat{\cal H}\vert{3,l}\rangle=(E_\text{FM}-2J_1-J_2+h)\vert{3,l}\rangle + J_2\vert{2,l}\rangle \nonumber
\\
&& \qquad\qquad\quad +J_1\big(\vert{1,l}\rangle +\vert{1,l+1}\rangle\big).
\end{eqnarray}
By translational invariance, the probabilities $|c_{j,l}|^2$ should be independent of the cell index $l$, and hence, the probability amplitudes have the form of plane waves $c_{j,l} = f_j {\rm e}^{{\rm i} k l}$. The eigenvalue problem of the Hamiltonian (\ref{Eq:hamiltonian}) in a one-magnon subspace therefore reduces to a homogeneous system of three linear equations for the coefficients $f_1, f_2$ and $f_3$:
\begin{equation}
\!\!
\begin{pmatrix} \!
 -4J_1\!+\! h\!-\!\epsilon & J_1(1\!+\!{\rm e}^{-{\rm i}k}) & J_1(1\!+\!{\rm e}^{-{\rm i}k}) \!\!\! \\ \!
 J_1(1\!+\!{\rm e}^{{\rm i}k}) & -2J_1\!-\!J_2\!+\!h\!-\!\epsilon & J_2 \!\!\! \\ \!
 J_1(1\!+\!{\rm e}^{{\rm i}k}) & J_2 & -2J_1\!-\!J_2\!+\!h\!-\!\epsilon \!\!
    \end{pmatrix} \!\!\!
 \begin{pmatrix}
    f_1 \\
    f_2 \\
    f_3
 \end{pmatrix}\!\!=\!0, \nonumber
  \label{Eq:e_matrix}
\end{equation}
where $\epsilon= E- E_\text{FM}$ denotes the energy of one-magnon eigenstate relative to the fully polarized FM state. Solving the corresponding characteristic equation yields three distinct one-magnon branches:
\begin{eqnarray}\label{Eq:Epsilons}
&&\varepsilon_1 = h-2J_1-2J_2, \nonumber \\
&&\varepsilon_2 = h-J_1(3+\sqrt{5+4\cos k}), \nonumber \\
&&\varepsilon_3 = h-J_1(3-\sqrt{5+4\cos k}).
\end{eqnarray}
Fig.~\ref{fig:FlatBand_epsilon} shows dispersion relations of the three one-magnon bands (\ref{Eq:Epsilons}) at zero field $h=0$ for three representative values of the interaction ratio $J_2/J_1$. Notably, the flat band corresponds to a magnon bound on the vertical dimer ($J_2$-bond) as given by the eigenvector $\frac{1}{\sqrt{2}}(\vert \!+\!1\rangle_{2,l}\vert 0\rangle_{3,l} - \vert 0\rangle_{2,l}\vert \!+\!1\rangle_{3,l})$. This dispersionless mode with the energy cost $\varepsilon_1=-2J_1-2J_2+h$ becomes lowest among all three branches in the frustrated region $J_2 > 2J_1$, whereby it is energetically more favorable with respect to the polarized triplet state (i.e. $\varepsilon_1<0$) for the magnetic fields $h<2J_1+2J_2$. In the highly frustrated regime $J_2 > 2J_1$ and magnetic fields $h<2J_1+2J_2$, the exact ground state of the spin-1 Heisenberg chain is the BMC with the energy $E_\text{BMC} = E_\text{FM} + N \varepsilon_1 = 2NJ_1 - NJ_2 - 2Nh$, which arises from crystallization of bound magnons on all vertical dimers as schematically  depicted in Fig. \ref{fig:SpinConfig}(b) and given by: 
\begin{figure}
\includegraphics[width=\columnwidth]{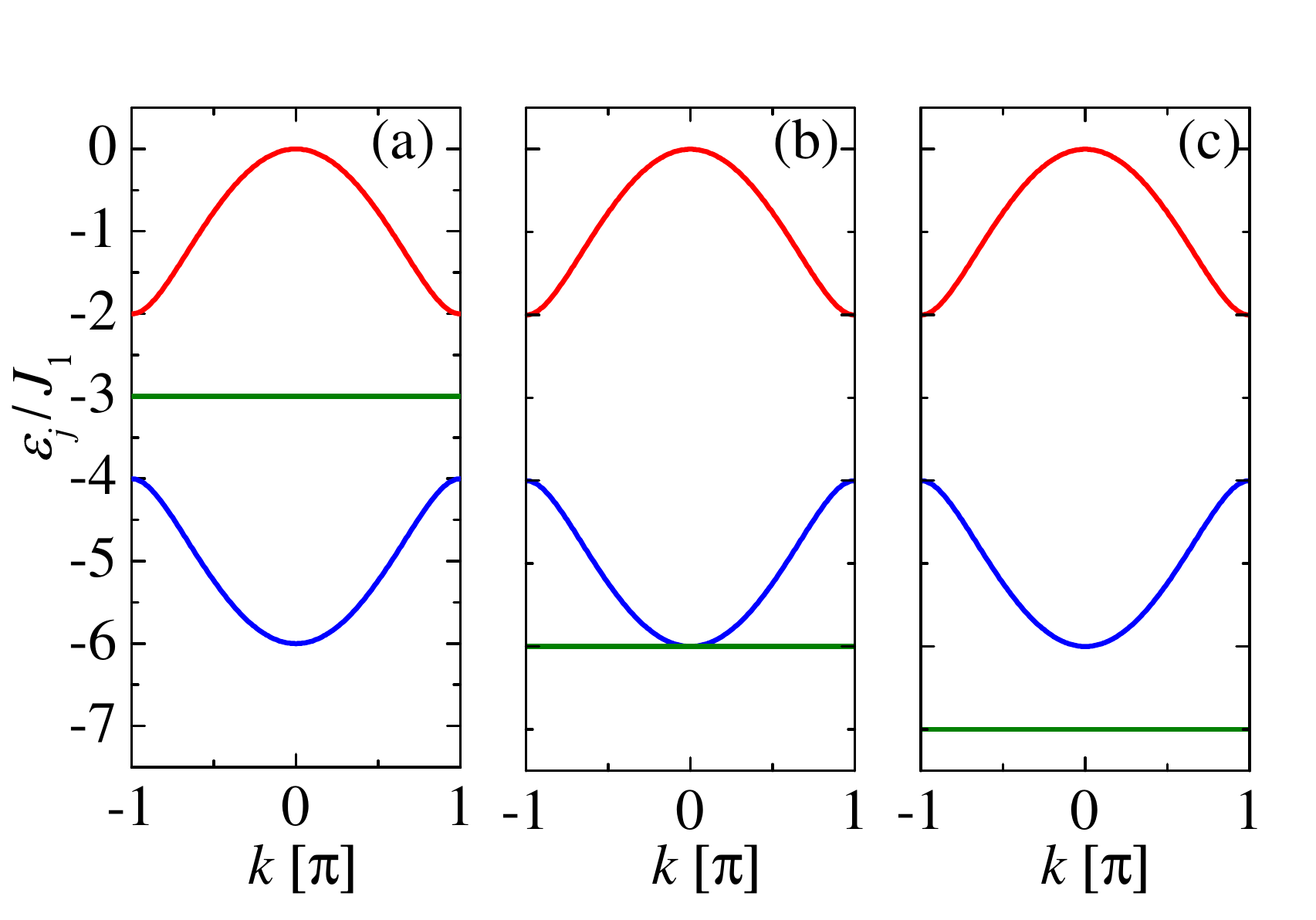}
\vspace{-0.9cm}
\caption{Three one-magnon energy bands of the spin-1 Heisenberg diamond chain as given by Eq. (\ref{Eq:Epsilons}) at zero magnetic field $h=0$ and three selected values of the interaction ratio: (a) $J_2/J_1=0.5$; (b) $J_2/J_1=2$; (c) $J_2/J_1=2.5$.}
\label{fig:FlatBand_epsilon}
\end{figure}
\begin{eqnarray}\label{Eq:BMC}
\vert\text{BMC}\rangle=\prod_{i=1}^{N}\vert \!+\!1\rangle_{1,l}\dfrac{1}{\sqrt{2}}\big(\vert \!+\!1\rangle_{2,l}\vert 0\rangle_{3,l} - \vert 0\rangle_{2,l}\vert \!+\!1\rangle_{3,l}\big). \nonumber \\
\end{eqnarray}
In the frustrated regime $J_2>2J_1$, the BMC phase becomes the ground state only when its energy is simultaneously lower than that of MD phase, i.e. $E_\text{BMC} < E_\text{MD}$, which limits its stability to a range of moderately strong magnetic fields $2J_1+J_2<h<2J_1+2J_2$. 

In the complementary parameter space $J_2 < 2J_1$, the lowest one-magnon eigenstate instead originates from the dispersive band $\varepsilon_2 = h-J_1(3+\sqrt{5+4\cos k})$ with a minimum $\varepsilon_2 = h - 6J_1$ at the wave number $k=0$, which corresponds to a low-momentum magnon delocalized over the entire spin-1 diamond chain. Consequently, the collective one-magnon state of the full spin-1 Heisenberg diamond chain becomes the respective ground state just below the saturation field $h=6J_1$ in the parameter region $J_2 < 2J_1$.

\subsection{Density-matrix renormalization group method} 

Although the spin-1 Heisenberg diamond chain is not integrable, it still belongs to a special class of frustrated quantum spin models with locally conserved quantities, because its Hamiltonian (\ref{Eq:hamiltonian}) commutes with the composite spin operator of the $i$-th dimer $[\hat{\cal H}, \boldsymbol{\hat{S}}_{23,i}^{2}] = 0$. This property allows the Hamiltonian to be conveniently rewritten in terms of the composite spin operator of the $i$-th dimer $\boldsymbol{\hat{S}}_{23,i}$: 
\begin{eqnarray}
\hat{\cal H} = J_1 \! \sum_{i=1}^{N} \! (\boldsymbol{\hat{S}}_{1,i} \!+\! \boldsymbol{\hat{S}}_{1,i+1}) \!\cdot\! \boldsymbol{\hat{S}}_{23, i} 
+ \frac{J_2}{2} \! \sum_{i=1}^{N} \! (\boldsymbol{\hat{S}}_{23, i}^2 \!-\! 4).
\label{hamlcl}
\end{eqnarray}
The Zeeman's term is omitted here as it only produces a trivial shift of the respective energy eigenvalues. Similarly, the latter part of the effective Hamiltonian (\ref{hamlcl}) also merely shifts the energy eigenvalues depending on specific values of the composite quantum spin numbers achieving one of three available values $S_{23,i} = 0$, $1$ or $2$. By contrast, the former part of the effective Hamiltonian (\ref{hamlcl}) corresponds to mixed spin-(1, $S_{23,i}$) Heisenberg chains, which can be efficiently solved numerically using the density-matrix renormalization group (DMRG) method. Our DMRG simulations were carried out by adapting the routine from the Algorithms and Libraries for Physics Simulations (ALPS) project \cite{baue11} for the effective mixed spin-(1, $S_{23,i}$) Heisenberg chains with $N=60$ unit cells, which correspond to a total number of $180$ spins in the original spin-1 Heisenberg diamond chain.  

The ground state of the spin-1 Heisenberg diamond chain can be consequently found from the lowest-energy eigenstates of the effective Hamiltonian (\ref{hamlcl}) by considering all possible combinations of the composite quantum spin numbers $S_{23,i}$. If translational period of the magnetic ground state is not broken, the corresponding lowest-energy eigenvalues of the effective Hamiltonian (\ref{hamlcl}) of the spin-1 Heisenberg diamond chain are obtained from the following formulas:
\begin{eqnarray}
&& E_{0} (2N, S_T^z = N) = -2 N J_2, \label{ee10} \\
&& E_{1} (2N, S_T^z) = 2 N J_1 \varepsilon_{1,1} (2N, S_T^z) - N J_2, \label{ee11} \\
&& E_{2} (2N, S_T^z) = 2 N J_1 \varepsilon_{1,2} (2N, S_T^z) + N J_2. \label{ee12}
\end{eqnarray}
where $\varepsilon_{1,1} (2N, S_T^z)$ and $\varepsilon_{1,2} (2N, S_T^z)$ denote the lowest-energy eigenvalues per spin of the antiferromagnetic spin-({\green 1}, {\orange 1}) Heisenberg chain and the ferrimagnetic mixed spin-({\green 1}, {\orange 2}) Heisenberg chain with unit coupling constant and the total number of $2N$ spins in the sector with the $z$-component of the total spin $S_T^z$. The lowest-energy eigenvalue $E_{0} (2N, S_T^z=N) = -2 N J_2$ apparently corresponds to the exact MD ground state (\ref{Eq:M-D}) depicted in Fig. \ref{fig:SpinConfig}(a) with the paramagnetic character of the monomeric spins and all dimers in the singlet state $S_{23,i} = 0$ in agreement with the previous variational analysis. When all vertical dimers are in a triplet state $S_{23,i} = 1$, the effective Hamiltonian (\ref{hamlcl}) reduces to the antiferromagnetic spin-({\green1}, {\orange 1}) Heisenberg chain, which has three available ground states - the gapped Haldane phase (the sector $S_T^z=0$), the gapless Tomonaga-Luttinger quantum spin liquid QSL-II (the sectors $S_T^z=1, \ldots, 2N-1$), and the fully polarized state (the sector $S_T^z=2N$) at low, moderate, and high magnetic fields, respectively. The fully polarized state of the effective spin-({\green 1}, {\orange 1}) Heisenberg antiferromagnetic chain with the energy $E_{1} (2N, S_T^z=2N) = 2NJ_1 - N J_2$ coincides with the exact BMC ground state (\ref{Eq:BMC}) depicted in Fig. \ref{fig:SpinConfig}(b), which was previously identified within the concept of localized magnons. Furthermore, the ferrimagnetic mixed spin-({\green 1}, {\orange 2})  Heisenberg chain is obtained from the effective Hamiltonian (\ref{hamlcl}) if all vertical dimers are in the quintet state $S_{23,i} = 2$, which favors the Lieb-Mattis ferrimagnetic ground state FRI-I (the sector with $S_T^z=N$) at low magnetic fields and other Tomonaga-Luttinger quantum spin liquid QSL-I (the sectors $S_T^z=N+1, \ldots, 3N-1$) at higher magnetic fields. 

However, the magnetic ground state of the spin-1 Heisenberg diamond chain may also spontaneously break translational symmetry, in which case the composite quantum spin number $S_{23,i}$ associated with spin dimers on the vertical $J_2$-bonds does not remain uniform along the entire chain. To account for this possibility, we examined higher-period magnetic eigenstates of the effective mixed spin-$(1, S_{23,i})$ Heisenberg chain considering translational periods up to four unit cells. This analysis revealed an additional ferrimagnetic ground state FRI-II of the spin-1 Heisenberg diamond chain, which arises from a mixed spin-({\green 1}, {\orange 1}, {\green 1}, {\orange 2}) Heisenberg chain exhibiting period doubling due to the regular alternation of dimer-triplet and dimer-quintet states with the following energy:
\begin{eqnarray}
E_{1-2} (2N, S_T^z \!=\! \tfrac{N}{2}) = 2 N J_1 \varepsilon_{1, 1, 1, 2} (2N, S_T^z \!=\! \tfrac{N}{2}). 
\label{eeeffd}
\end{eqnarray}
Here, $\varepsilon_{1, 1, 1, 2} (2N, S_T^z = \tfrac{N}{2})$ denotes the lowest-energy eigenvalue per spin of the mixed-spin Heisenberg chain composed of a regularly alternating spin-1 trimer unit and spin-2 entity, the unit coupling constant, and the total number of $2N$ spins in the specific sector with the $z$-component of the total spin $S_T^z = \tfrac{N}{2}$. 

\subsection{Lanczos diagonalization}

\begin{figure}
  \centering
  \resizebox{0.5\textwidth}{!}{
 \includegraphics[trim = 50 250 50 80, clip]{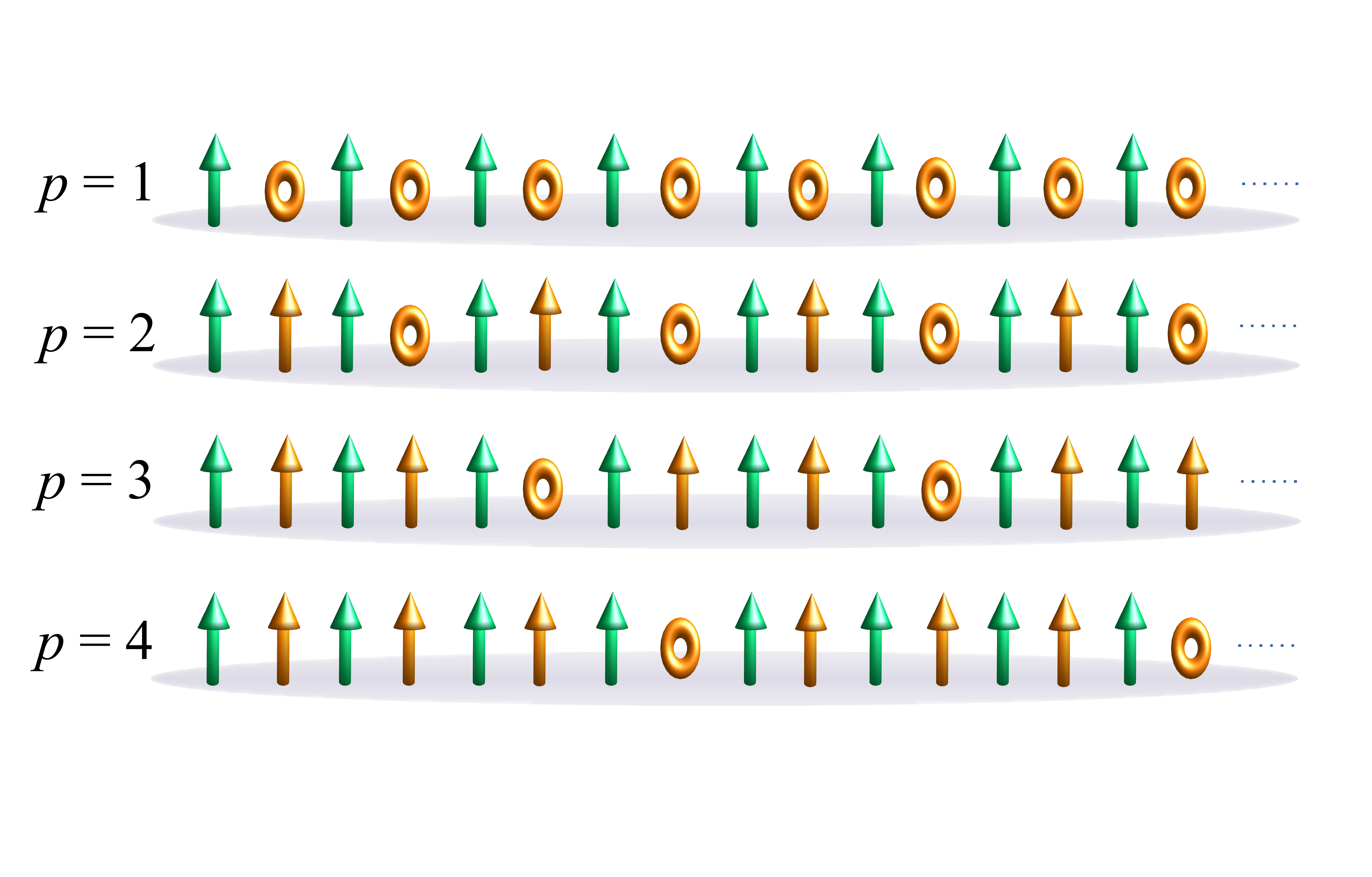}}
\vspace{-0.7cm}
\caption{A schematic illustration of all fragmented ground states of the spin-1 Heisenberg diamond chain. Except the fully fragmented MD phase with unique singlet state of all dimers corresponding to a magnetic ground state with the period 
$p = 1$, three fragmented cluster-based Haldane phases denoted as the tetramer-dimer (TD) phase with the period $p = 2$, the heptamer-dimer (HD) phase with the period $p = 3$, and the decamer-dimer (DD) phase with the period $p = 4$ emerge. Orange arrows and zeros represent dimer-triplet and dimer-singlet states of the vertical spin-1 dimers, respectively, while green arrows correspond to the monomer spins.}
\label{fig:HaldaneSpinConfig}
\end{figure}

On the other hand, the formation of the dimer-singlet state on the $J_2$-bonds fragments the effective mixed spin-(1, $S_{23,i}$) Heisenberg chains into smaller segments, which can be solved exactly using the Lanczos diagonalization method implemented within the ALPS project \cite{baue11}. This analysis reveals three distinct higher-period ground states with periods of two, three, and four unit cells arising from a regular alternation of dimer-triplet and dimer-singlet states. These phases emergent between the fully fragmented MD phase and the Haldane phase correspond to the lowest-energy eigenstates of the effective mixed spin-({\green 1}, {\orange 1}, {\green 1}, {\orange 0}), spin-({\green 1}, {\orange 1}, {\green 1}, {\orange 1}, {\green 1}, {\orange 0}) and spin-({\green 1}, {\orange 1}, {\green 1}, {\orange 1}, {\green 1}, {\orange 1}, {\green 1}, {\orange 0}) Heisenberg chains, respectively. The resulting higher-period ground states can be interpreted as cluster-based Haldane phases illustrated in Fig. \ref{fig:SpinConfig}(c)-(e) and Fig. \ref{fig:HaldaneSpinConfig} with energy eigenvalues given by:
\begin{eqnarray}
&& E_{1-0} (2N, S_T^z = \tfrac{N}{2}) = \frac{3N}{2} J_1 \varepsilon_{1} (3, S_T^z=1) - \frac{3N}{2} J_2, \nonumber \\
&& \hspace{3.1cm} = - \frac{3N}{2} J_1 - \frac{3N}{2} J_2, \nonumber \\
&& E_{1-1-0} (2N, S_T^z = \tfrac{N}{3}) = \frac{5N}{3} J_1 \varepsilon_{1} (5, S_T^z=1) - \frac{4N}{3} J_2, \nonumber \\
&& E_{1-1-1-0} (2N, S_T^z = \tfrac{N}{4}) = \frac{7N}{4} J_1 \varepsilon_{1} (7, S_T^z=1) - \frac{5N}{4} J_2. \nonumber \\
\label{eeeffd}
\end{eqnarray}    
Here, $\varepsilon_{1} (2n+1, S_T^z=1)$ denotes the lowest-energy eigenvalue of a finite spin-1 Heisenberg antiferromagnetic chain with an odd number of spins $2n+1$ and unit coupling constant, which always belongs to the triplet sector $S_T^z=1$ due to open boundary conditions imposed by two adjacent dimer-singlet states. The fragmented ground states with periods $p=2$, $3$, and $4$ illustrated in Fig. \ref{fig:SpinConfig}(c)-(e) and Fig. \ref{fig:HaldaneSpinConfig} correspond to the cluster-based Haldane phases with character of the tetramer-dimer (TD) phase, heptamer-dimer (HD) phase, and decamer-dimer (DD) phase, respectively.

\subsection{Ground-state phase diagram}

\begin{figure}
\centering
\includegraphics[width=0.5\textwidth]{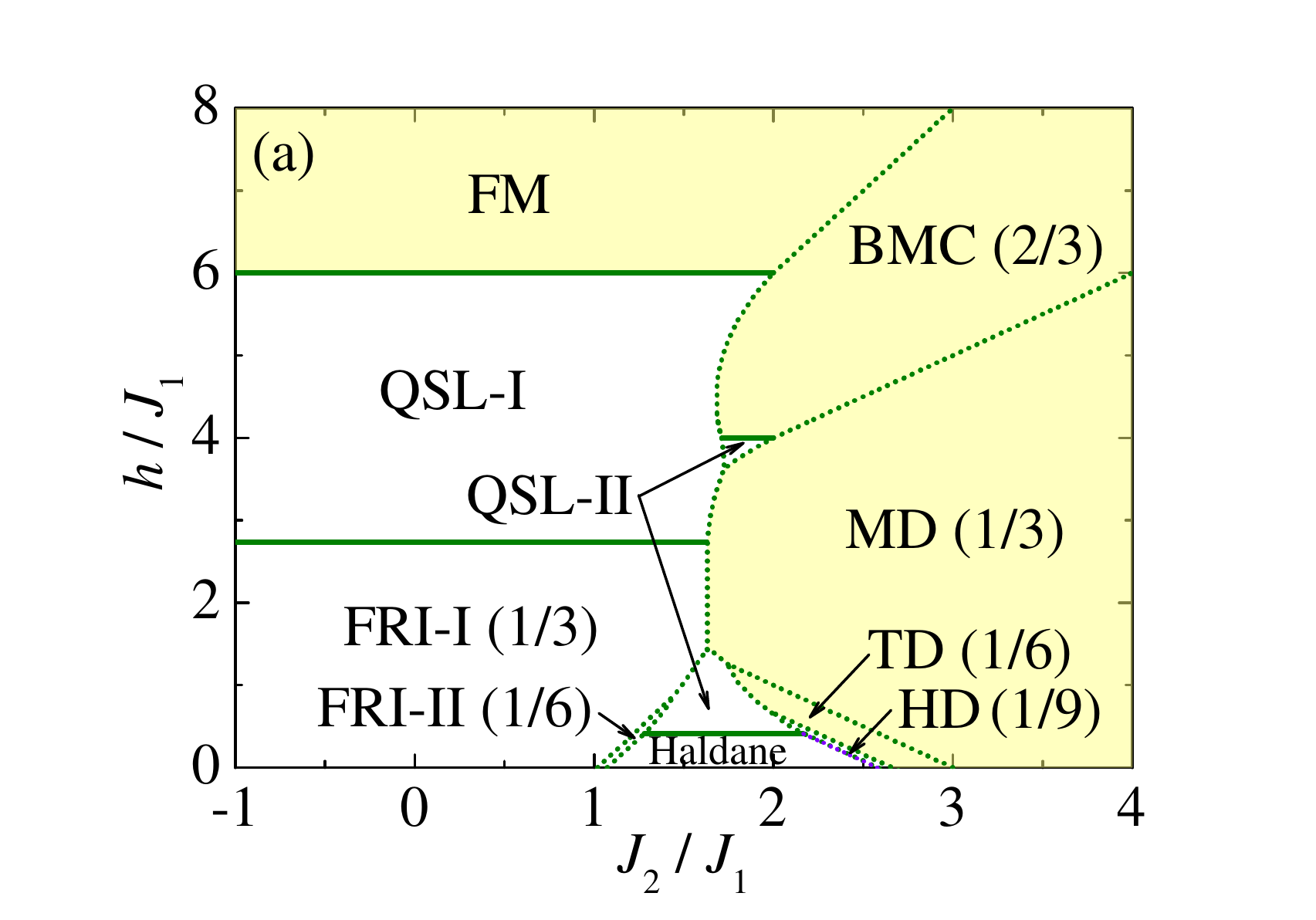}
\includegraphics[width=0.5\textwidth]{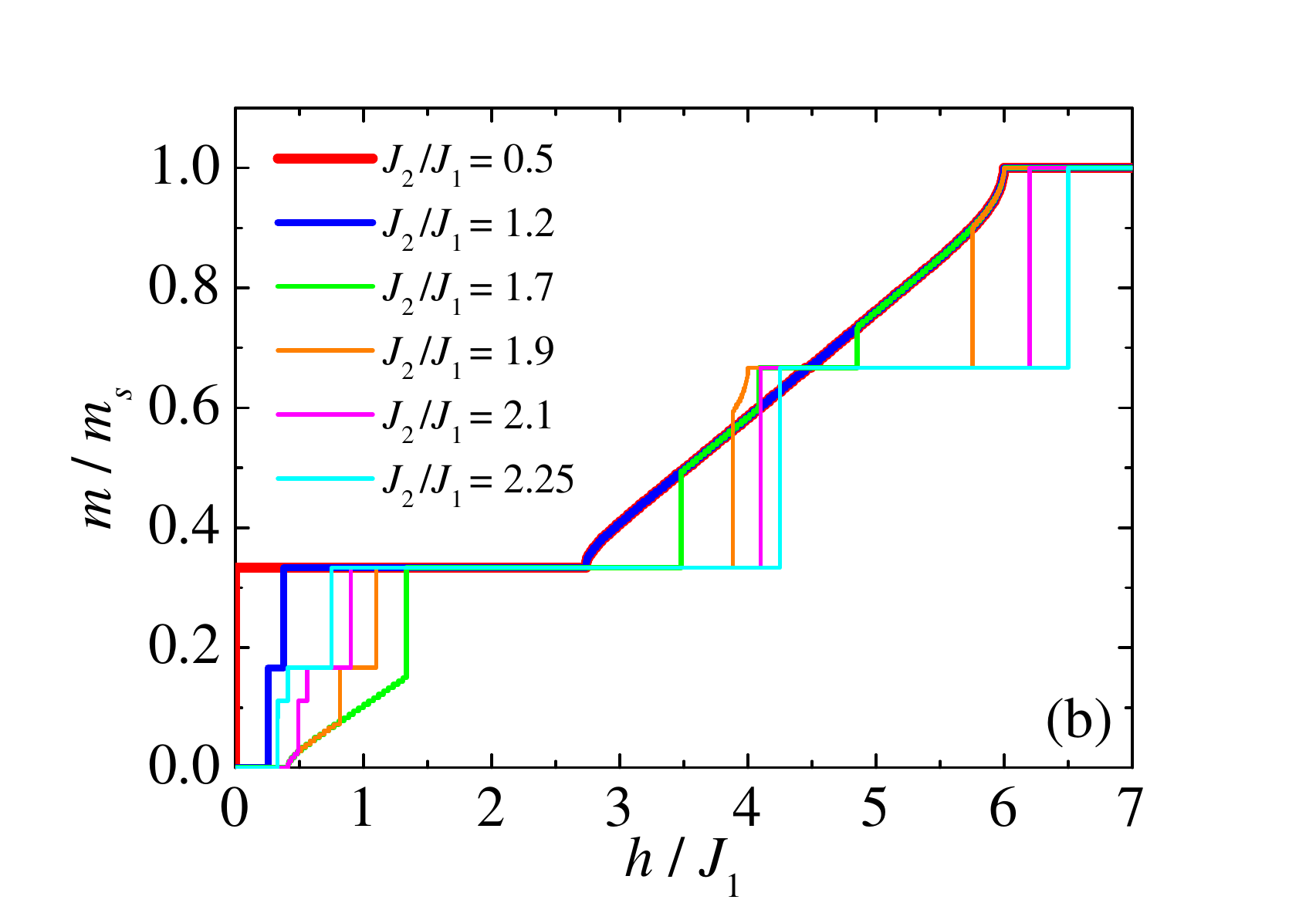} 
\vspace{-0.9cm}
\caption{(a) Ground-state phase diagram of the spin-1 Heisenberg diamond chain in the $J_2/J_1-h/J_1$ plane. Continuous (discontinuous) lines mark continuous (discontinuous) quantum phase transitions. The notation for individual ground states: the ferrimagnetic phases (FRI-I and FRI-II), the quantum spin-liquid phases (QSL-I and QSL-II), the monomer-dimer (MD) phase, the cluster-based Haldane phases - tetramer-dimer (TD) and heptamer-dimer (HD) phase, the bound-magnon crystal (BMC) phase, the saturated ferromagnetic (FM) phase. Numbers in parentheses indicate magnetization normalized to the saturation value. Violet dots mark the narrow region hosting the decamer-dimer (DD) cluster-based Haldane phase; (b) Several representative zero-temperature magnetization curves of the spin-1 Heisenberg diamond chain obtained from DMRG simulations.}
\label{fig:GSPD}
\end{figure}

The exact analytical and numerical results are now combined to construct the complete ground-state phase diagram of the spin-1 Heisenberg diamond chain, which is depicted in the $J_2/J_1-h/J_1$ plane in Fig. \ref{fig:GSPD}(a). The diagram is rather complex and comprises two ferrimagnetic ground states FRI-I and FRI-II corresponding to intermediate magnetization plateaus at one-third and one-sixth of the saturation magnetization, two distinct gapless Tomonaga-Luttinger quantum spin-liquid phases QSL-I and QSL-II with continuously varying magnetization, the fully fragmented MD phase corresponding to another type of intermediate one-third magnetization plateau, and three fragmented cluster-based Haldane phases TD, HD, and DD associated with one-sixth, one-ninth, and one-twelfth magnetization plateaus. In addition, the diagram contains the trivial fully polarized FM phase, the BMC ground state featuring an intermediate two-thirds magnetization plateau, and the uniform gapped Haldane phase with zero magnetization. All horizontal boundaries represent continuous field-driven quantum phase transitions, whereas discontinuous magnetization jumps occur across all other phase boundaries with nonzero slope. Notably, these results are fully consistent with earlier findings reported by Hida and Takano for the zero-field case of the spin-1 Heisenberg diamond chain \cite{hida17} further validating our comprehensive phase classification. 

To verify the validity of the established ground-state phase diagram, we present in Fig. \ref{fig:GSPD}(b) several representative zero-temperature magnetization curves. Magnetization plateaus at fractional values of the saturation magnetization are indicative of distinct quantum ground states with a finite energy gap discussed in the preceding ground-state analysis, whereas abrupt magnetization jumps signal discontinuous magnetic-field-driven phase transitions between them. In contrast, the smooth and continuous increase of zero-temperature magnetization with the magnetic field contrarily reflects the presence of Tomonaga-Luttinger spin liquid phases QSL-I and QSL-II characterized by the absence of an energy gap. 

\section{Results and discussion}
\label{FTM}

To bring insight into the finite-temperature behavior of the spin-1 Heisenberg diamond chain, we employed several complementary computational schemes including ED, QMC, and generalized localized-magnon theory. Full ED calculations were carried out for spin-1 Heisenberg diamond chains up to 12 spins ($N = 4$ unit cells) using the fulldiag routine from the open-source ALPS library \cite{baue11} by imposing periodic boundary conditions. In the unfrustrated regime, QMC simulations were alternatively performed for the spin-1 Heisenberg diamond chains for much larger system sizes up to 180 spins ($N = 60$ unit cells) using the stochastic-series expansion routine dirloopsse implemented within the ALPS project \cite{baue11}. For the frustrated regime hosting fragmented ground states and bound magnons, we developed an effective monomer-dimer lattice-gas model formulated within the extended localized-magnon theory to examine the magneto-thermodynamics of the spin-1 Heisenberg diamond chain at finite temperatures \cite{karl19,karl22,stre22}.

\begin{figure}
\centering
\includegraphics[width=0.5\textwidth]{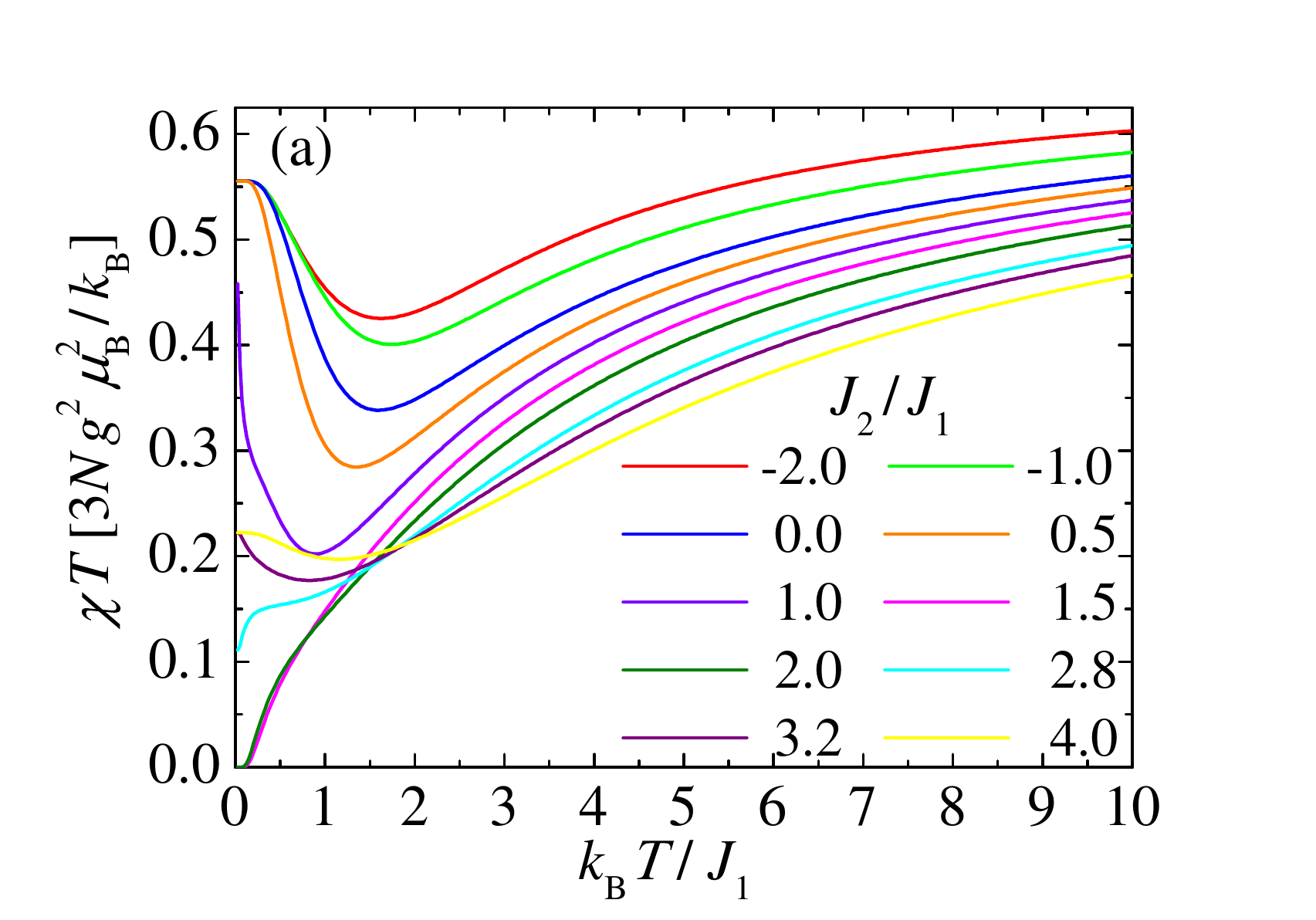}
\vspace*{-0.4cm}
\includegraphics[width=0.5\textwidth]{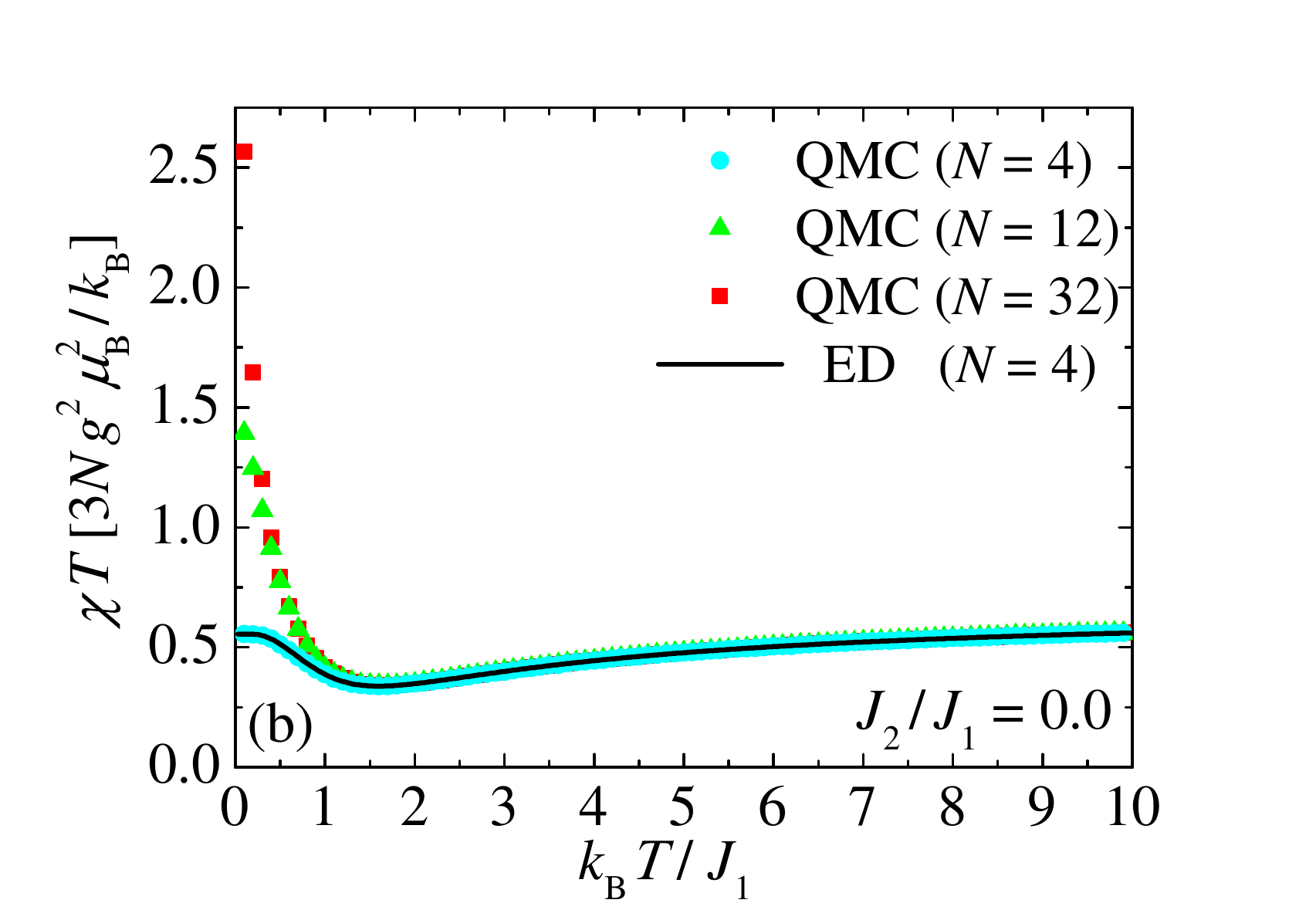}
\vspace{-0.5cm}
\caption{(a) Temperature variations of the magnetic susceptibility times temperature product ($\chi T$) of the spin-1 Heisenberg diamond chain in zero magnetic field for a few selected values of the interaction ratio $J_2/J_1$. The data were obtained from the full ED of a finite-size chain with 12 spins and are normalized per spin; (b) Comparison of full ED data for the finite-size chain with 12 spins ($N=4$) with QMC simulations of the finite-size chains with 12, 36, and 96 spins ($N=4$, 12, 32) for the particular value of the interaction ratio $J_2/J_1=0$.}
\label{fig:chit}
\end{figure}

\subsection{Magnetic susceptibility}
\label{MS}
We begin by analyzing typical temperature variations of the magnetic susceptibility times temperature ($\chi T$) product for the spin-1 Heisenberg diamond chain in zero magnetic field. Calculations were performed using full ED of a finite chain of 12 spins ($N = 4$ unit cells) for a few selected values of the interaction ratio as the $\chi T$ product often serves as a primary benchmark for interpreting experimental magnetic behavior. In the high-temperature limit, the $\chi T$ product converges to the asymptotic value 2/3 consistent with the Curie constant of the spin-1 entities independently of the interaction ratio. The thermal behavior at low and moderate temperatures basically depends on the interaction ratio as illustrated in Fig. \ref{fig:chit}(a). It is therefore instructive to find out how fundamental differences between individual zero-field ground states are manifested in the respective thermal dependencies of the $\chi T$ product. For sufficiently small values of the interaction ratio $J_2/J_1 < 1.02$, the $\chi T$ product exhibits upon cooling the typical signatures of quantum ferrimagnets: it initially shows a gradual decline to a local minimum before it starts to rise towards the zero-temperature asymptotic value of 5/9. This value is consistent with the ferrimagnetic character of the LM ground state FRI-I belonging to the sector $S_T^z = 4$ when considering the finite diamond chain with $N = 4$ unit cells. Note furthermore that the thermal dependence for the particular ratio $J_2/J_1 = 1.0$ apparently shows a very steep upturn to the initial zero-temperature asymptotic value 5/9. This low-temperature enhancement originates from low-lying excitations into another quantum ferrimagnetic phase FRI-II, which becomes the ground state in a narrow interval of the interaction ratio $J_2/J_1 \in(1.02, 1.07)$. The gapped Haldane ground state, which is realized over a broader range of the interaction parameters $J_2/J_1 \in(1.07, 2.58)$, manifest itself through a markedly different thermal dependence. Indeed, the $\chi T$ product monotonically rises with increasing of the temperature from zero initial value consistent with the nonmagnetic character of this zero-field ground state. On the contrary, the cluster-based Haldane phases HD and TD realized as the respective zero-field ground states in the parameter space $J_2/J_1 \in(2.58, 2.66)$ and $J_2/J_1 \in(2.66, 3.0)$ have nonzero net magnetic moment $S_T^z=1$ and $S_T^z=2$ for a finite-size chain of $N = 4$ unit cells due to a triplet character of the heptamers and tetramers, respectively [see Fig. \ref{fig:SpinConfig}(c) and (d)]. The particular case $J_2/J_1 = 2.8$ for instance illustrates how the $\chi T$ product asymptotically tends towards value $1/9$ in zero-temperature limit reflecting the total spin moment $S_T^z=2$ of the TD ground state. Last but not least, the zero-temperature asymptotic limit of the $\chi T$ product reaches the value $2/9$ in the highly frustrated regime $J_2/J_1>3$ because of paramagnetic contribution of four monomeric spins in the MD ground state for a finite chain with $N = 4$ unit cells. In the thermodynamic limit $N \to \infty$, one should observe for all magnetic ground states with nonzero net spin moment zero-temperature divergence of the $\chi T$ product instead of approaching finite values, which are mere artefacts of finite-size calculations. Aside from this low-temperature limitation, the results correctly reproduce general trends in the thermal dependence of the $\chi T$ product as evidenced by the comparison of full ED data (symbols) for a finite chain with 12 spins and QMC simulations for significantly larger system sizes with 12, 36, and 96 spins as shown in Fig. \ref{fig:chit}(b).

\begin{figure}
\centering
\includegraphics[width=0.5\textwidth]{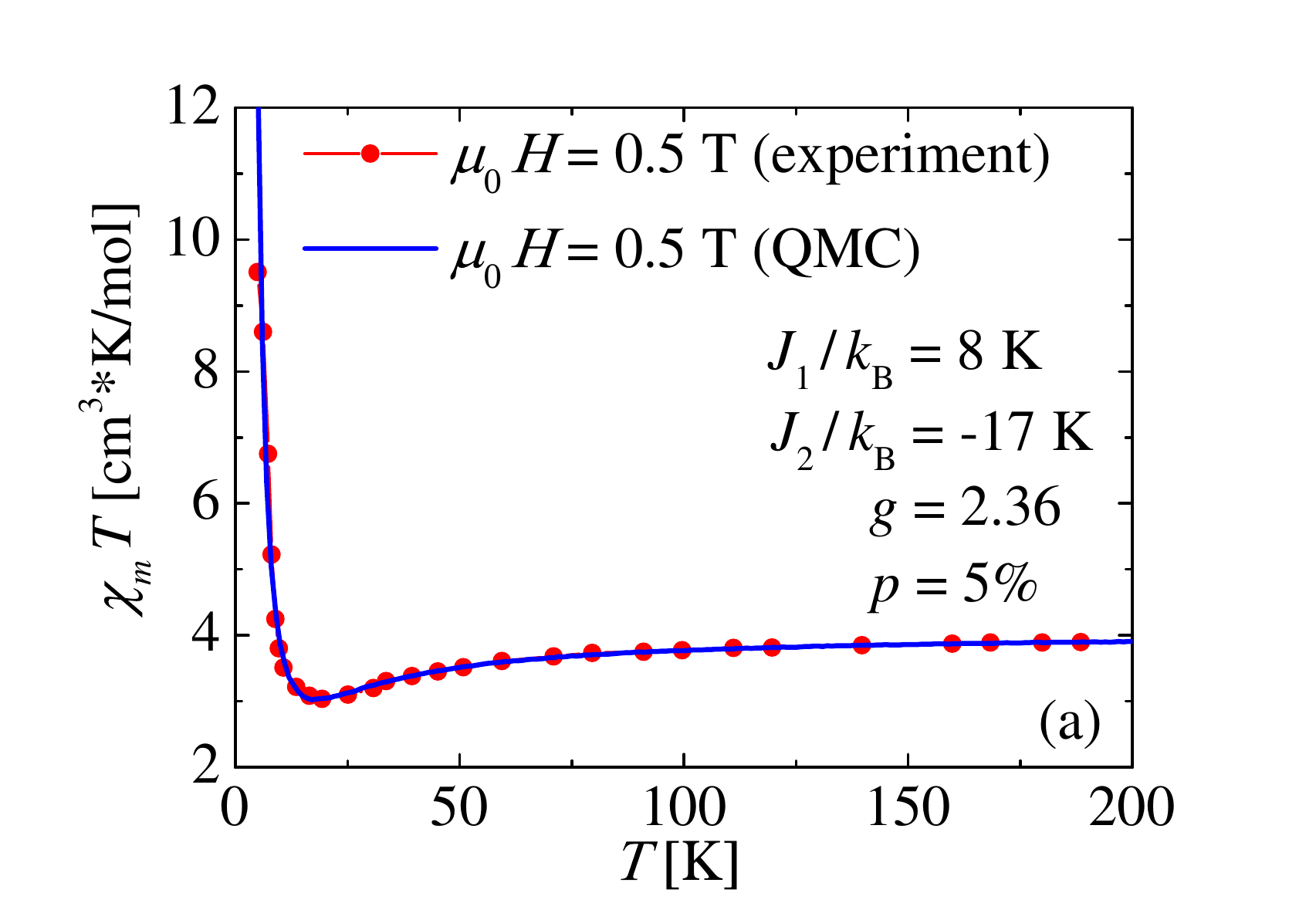}
\vspace*{-0.4cm}
\includegraphics[width=0.5\textwidth]{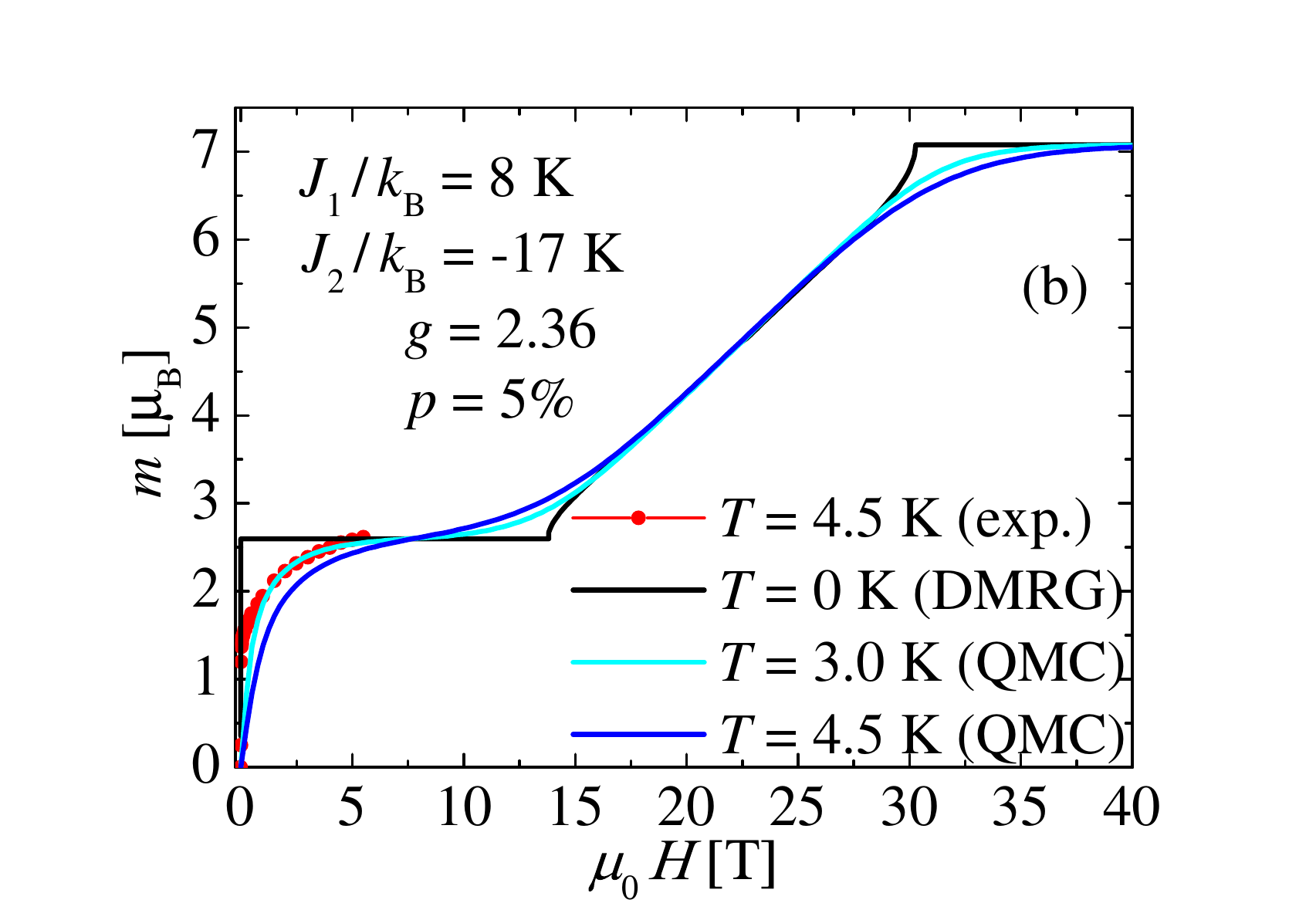}
\vspace*{-0.4cm}
\caption{(a) Temperature variations of the molar susceptibility times temperature product ($\chi_m T$) of the polymeric compound [Ni$_3$(OH)$_2$(C$_4$H$_2$O$_4$)(H$_2$O)$_4$] $\cdot$ 2H$_2$O (Ni$_3$-dc) together with the best theoretical fit obtained from QMC simulations of the spin-1 Heisenberg diamond chain with 180 spins; (b) Field dependence of the magnetization of the polymeric compound Ni$_3$-dc measured at $T=4.5$~K compared with theoretical data for the spin-1 Heisenberg diamond chain with 180 spins at $T=0$~K (DMRG), $T=3$~K (QMC), and $T = 4.5$~K (QMC). Experimental data for the molar susceptibility and magnetization both normalized per formula unit were taken from Ref. \cite{guil02}, while the best simultaneous fit of the susceptibility and magnetization data was obtained for the coupling constants $J_1/k_{\rm B} = 8$~K and $J_2/k_{\rm B} = -17$~K, Land\'e g-factor $g=2.36$ and $5$\% of Ni$^{2+}$ paramagnetic impurities.}
\label{fig:exp}
\end{figure}

\subsection{Ni$_3$-dc compound: theoretical interperation}
\label{TVE}

Next, the spin-1 Heisenberg diamond chain will be used to interpret available experimental data for the magnetic susceptibility and magnetization of the polymeric coordination compound [Ni$_3$(OH)$_2$(C$_4$H$_2$O$_4$)(H$_2$O)$_4$] $\cdot$ 2H$_2$O hereafter abbreviated as Ni$_3$-dc. Upon lowering of temperature, the $\chi_m T$ product of the Ni$_3$-dc decreases from its high-temperature value $3.89$~emu$\cdot$K/mol to a flat minimum of $3.03$~emu$\cdot$K/mol near 20~K followed by a pronounced low-temperature upturn indicative of zero-temperature divergence, see Fig.~\ref{fig:exp}(a). Moreover, the isothermal magnetization curve measured for the Ni$_3$-dc compound at $4.5$~K exhibits at low fields a sharp rise, which is subsequently followed by a weak-field dependence indicative of an intermediate one-third magnetization plateau. These observations are compatible either with magnetic features of the LM ferrimagnetic phase FRI-I or the fully fragmented MD phase. However, the latter MD phase fails to reproduce the position and depth of the flat minimum in the temperature dependence of $\chi_m T$ product. To capture an abrupt low-field upturn of the magnetization at $4.5$~K, we additionally incorporated a small amount of Ni$^{2+}$ paramagnetic impurities. Under this assumption, we obtained a reasonable simultaneous fit of both magnetic susceptibility and magnetization data of the coordination polymer Ni$_3$-dc refined by 5 \% of Ni$^{2+}$ paramagnetic impurities by performing QMC simulations of the spin-1 Heisenberg diamond chain with a dominant ferromagnetic intra-dimer coupling $J_2/k_{\rm B} = -17$~K and a weaker antiferromagnetic monomer-dimer coupling constant $J_1/k_{\rm B} = 8$~K. These values place Ni$_3$-dc compound to the parameter regime of the ferrimagnetic LM ground state FRI-I, which is expected to collapse at a field-induced quantum phase transition into the Tomonaga-Luttinger quantum spin liquid QSL-I emerging approximately at $14$~T. In light of the absence of high-field magnetization data above $5.5$~T, targeted high-field measurements would provide a decisive test of the predicted quantum phase transition into the quantum spin liquid QSL-I at $14$~T, as well as, the subsequent field-induced quantum phase transition from the quantum spin liquid QSL-I to the fully polarized FM phase near $30$~T.    

\subsection{Effective monomer-dimer lattice-gas description}
\label{LMT}

\begin{figure}[t]
\centering
\includegraphics[width=0.45\textwidth]{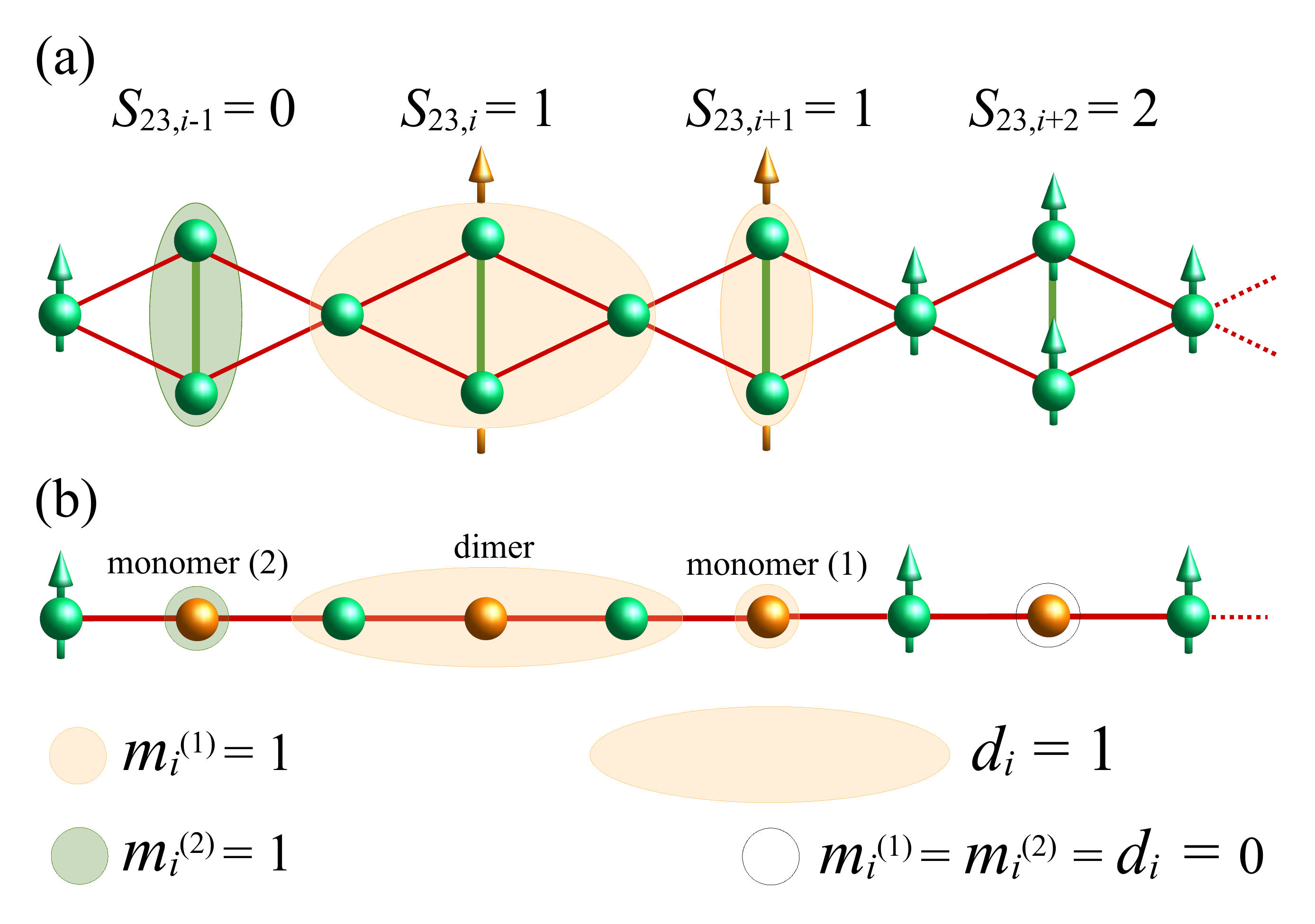} 
\vspace{-0.5cm}
\caption{A schematic illustration of a representative spin configuration of the spin-1 Heisenberg diamond chain [panel (a)] together with its corresponding quasi-particle representation within the effective monomer-dimer lattice-gas model [panel (b)]. The effective lattice-gas model is featuring two types of monomeric particles and one type of dimeric particles, which correspond to a bound one-magnon state ($S_{23,i}=1$, orange circle), a bound two-magnon state ($S_{23,i}=0$, green circle), a triplet state of a diamond plaquette ($S_{23,i}=1$, orange oval), or a fully polarized state ($S_{23,i}=2$, empty circle). }
\label{fig:LG_SpinConfig}
\end{figure}

In the highly frustrated regime highlighted in Fig. \ref{fig:FlatBand_epsilon} as the shaded area encompassing the TD, MD, BMC, and FM phases, we establish a mapping correspondence between the spin-1 Heisenberg diamond chain given by the Hamiltonian (\ref{Eq:hamiltonian}) and an effective monomer-dimer lattice-gas model providing a simple framework for evaluating its magnetic, magnetocaloric, and thermodynamic quantities \cite{karl19,karl22,stre22}. Considering bound one- and two-magnon eigenstates of the spin-1 dimers on the vertical $J_2$-bonds together with the lowest-energy triplet eigenstates of an elementary spin-1 diamond plaquette indeed captures the three distinct quantum ground states TD, MD, and BMC of the spin-1 Heisenberg diamond chain emerging in the frustrated parameter regime. Consequently, the spin-1 Heisenberg diamond chain can be effectively mapped onto a monomer-dimer lattice-gas model governed by the effective Hamiltonian:
\begin{eqnarray}\label{Eq:H_eff}
H_\text{eff}= E_\text{FM}  - \sum\limits_{i=1}^N \! \left(\mu_1^{(1)} m_i^{(1)} \! + \mu_1^{(2)}  m_i^{(2)} \! + \mu_2^{(k)} d_i \! \right)\!, 
\end{eqnarray} 
where $E_\text{FM}= N(4J_1 + J_2) - 3Nh$ is the energy of the fully polarized FM state serving as a reference (vacuum) state, $m_i^{(j)} = 0,1$ ($j=1,2$) denote the occupation numbers of two types of monomeric particles, and $d_i = 0,1$ represents the occupation number of dimeric particles.  The corresponding chemical potentials quantify the excitation energies of these quasi-particles: the chemical potential $\mu_1^{(1)} = 2J_1 + 2J_2 - h$ of the first kind of monomeric particle measures the energy cost of creating a bound one-magnon state on a vertical dimer, the chemical potential $\mu_1^{(2)} = 4J_1 + 3J_2 - 2h$ of the second kind of monomeric particle determines the energy cost associated with the formation of a bound two-magnon state (i.e. the dimer-singlet state) on a vertical dimer, and the chemical potentials $\mu_2^{(k)} = 7J_1 + 2J_2 - k h $ $(k=3-5)$ of the dimeric particles represent the energy cost related to the creation of one of the three lowest-energy triplet states on an elementary spin-1 diamond plaquette composed of a single vertical dimer and its two neighboring monomer spins (see Fig. \ref{fig:LG_SpinConfig} for a schematic representation of these states and their quasi-particle representations).

To extract the free energy of the system, we first define the partition function of the effective monomer-dimer lattice-gas model given by the Hamiltonian (\ref{Eq:H_eff}):  
\begin{widetext}
\begin{eqnarray}\label{Eq:Z_eff0}
{Z}_\text{eff}= {\rm e}^{-\beta E_\text{FM}} \! \sum\limits_{\{m\}}\sum\limits_{\{d\}} 
\prod_{i=1}^N (1 \!-\! m_i^{(1)} \! m_i^{(2)}) (1 \!-\! m_i^{(1)} \! d_i) (1 \!-\! m_i^{(2)} \! d_i) (1 \!-\! d_i d_{i+1}) 
\exp\Bigl(\beta \sum\limits_{j=1}^2 \! \mu_1^{(j)} m_i^{(j)} \Bigr) \! \sum\limits_{k=3}^5 \! \exp \! \left(\beta \mu_2^{(k)} d_i \right)\!\!,
\end{eqnarray} 
\end{widetext}
where $\beta = 1/(k_{\rm B} T)$, $k_{\rm B}$ is Boltzmann's constant, $T$ is the absolute temperature, and the summations $\sum_{\{m\}}$ and $\sum_{\{d\}}$ are carried out over the sets of occupation numbers $\{ m_i^{(j)} \} $ and $\{ d_i \}$ of all monomeric and dimeric particles, respectively. The projection operators $(1 \!-\! m_i^{(1)} \! m_i^{(2)}) (1 \!-\! m_i^{(1)} \! d_i) (1 \!-\! m_i^{(2)} \! d_i) (1 \!-\! d_i d_{i+1})$ prevent in the effective monomer-dimer lattice-gas model multiple occupancy of the same site by more than one particle as well as the occupancy of adjacent sites by the dimeric particles. After performing summation over the occupation numbers of all monomeric particles, the partition function of the effective monomer-dimer lattice-gas model can be reformulated in terms of the transfer-matrix formalism:
\begin{eqnarray}\label{Eq:Z_eff1}
{Z}_\text{eff}= \exp(-\beta E_\text{FM})  \sum\limits_{\{d\}}\prod_{i=1}^N \text{T}(d_i,d_{i+1}), 
\end{eqnarray} 
where the expression $ \text{T}(d_i,d_{i+1})$ represents the transfer matrix depending on the occupation numbers of the dimeric particles from two adjacent sites: 
\begin{eqnarray}
\text{T}(d_i\!,d_{i+1}) &=& \left(1 - d_i d_{i+1} \right) \! \left[1 + 2 d_i \! \cosh(\beta h) \right] {\rm e}^{\beta \mu_2^{(4)} d_i} \nonumber \\
&\times& \left[1 + (1 - d_i) \left({\rm e}^{\beta \mu_1^{(1)}} + {\rm e}^{\beta \mu_1^{(2)}}\right)\right].
\label{Eq:TrM_eff1}
\end{eqnarray}
The transfer matrix $\text{T}(d_i,d_{i+1})$ defined by Eq. (\ref{Eq:TrM_eff1}) can be recast into the following matrix representation:
\begin{equation}\label{Eq:TrM_eff2}
\text{T}(d_i,d_{i+1}) = \begin{pmatrix}
\Omega & \Omega \\
e^{\beta \mu_2^{(4)}}[1 + 2\cosh(\beta h)] & 0
\end{pmatrix},
\end{equation} 
where $\Omega= 1 + {\rm e}^{\beta \mu_1^{(1)}} + {\rm e}^{\beta \mu_1^{(2)}}$. Within the transfer-matrix method, the partition function of the effective monomer-dimer lattice-gas model can be obtained after performing consecutive summation over the occupation numbers of the dimeric particles from two eigenvalues $\lambda_{\pm}$ of the transfer matrix (\ref{Eq:TrM_eff2}):
\begin{eqnarray}\label{Eq:PF_eff}
{Z}_\text{eff}= {\rm e}^{-\beta E_\text{FM}}\big(\lambda_+^N + \lambda_-^N\big),
\end{eqnarray}
which can be straightforwardly obtained by its direct diagonalization giving:
\begin{eqnarray}\label{Eq:lamda+}
{\lambda}_{\pm}=\dfrac{1}{2}\left(\Omega \pm \sqrt{\Omega^2 +4\Omega {\rm e}^{\beta \mu_2^{(4)}}[1+2\cosh(\beta h)]}\right)\!.
\end{eqnarray}
Now, we can calculate from the partition function (\ref{Eq:PF_eff}) the Gibbs free energy of the effective monomer-dimer lattice-gas model:
\begin{eqnarray}\label{Eq:GFE_eff}
{G}_\text{eff}= -k_{\rm B} T \ln {Z}_\text{eff} = E_\text{FM} - k_{\rm B} T \ln \left(\lambda_+^N + \lambda_-^N \right)\!.
\end{eqnarray}
The formula for the Gibbs free energy of the effective monomer-dimer model (\ref{Eq:GFE_eff}) reduces in the thermodynamic limit \(N \rightarrow \infty\) to much simpler expression depending solely on the largest eigenvalue ${\lambda}_+$ of the transfer matrix:
\begin{eqnarray}\label{Eq:GFETL_eff}
{G}_\text{eff}= E_\text{FM} - N k_{\rm B} T \ln \lambda_+.
\end{eqnarray}
To derive explicit expressions for all basic magnetic and thermodynamic quantities, one can employ the standard relations for the magnetization, entropy and specific heat:
\begin{eqnarray}
M = - \left( \frac{\partial {G}_\text{eff}}{\partial h} \right)_{\! T}\!\!\!\!, 
S = - \left( \frac{\partial {G}_\text{eff}}{\partial T} \right)_{\! h}\!\!\!\!,  
C = -T \left( \frac{\partial^2 {G}_\text{eff}}{\partial T^2} \right)_{\! h}\!\!\!\!. \nonumber
\end{eqnarray}

\begin{figure*}  
	\includegraphics[width=0.5\textwidth]{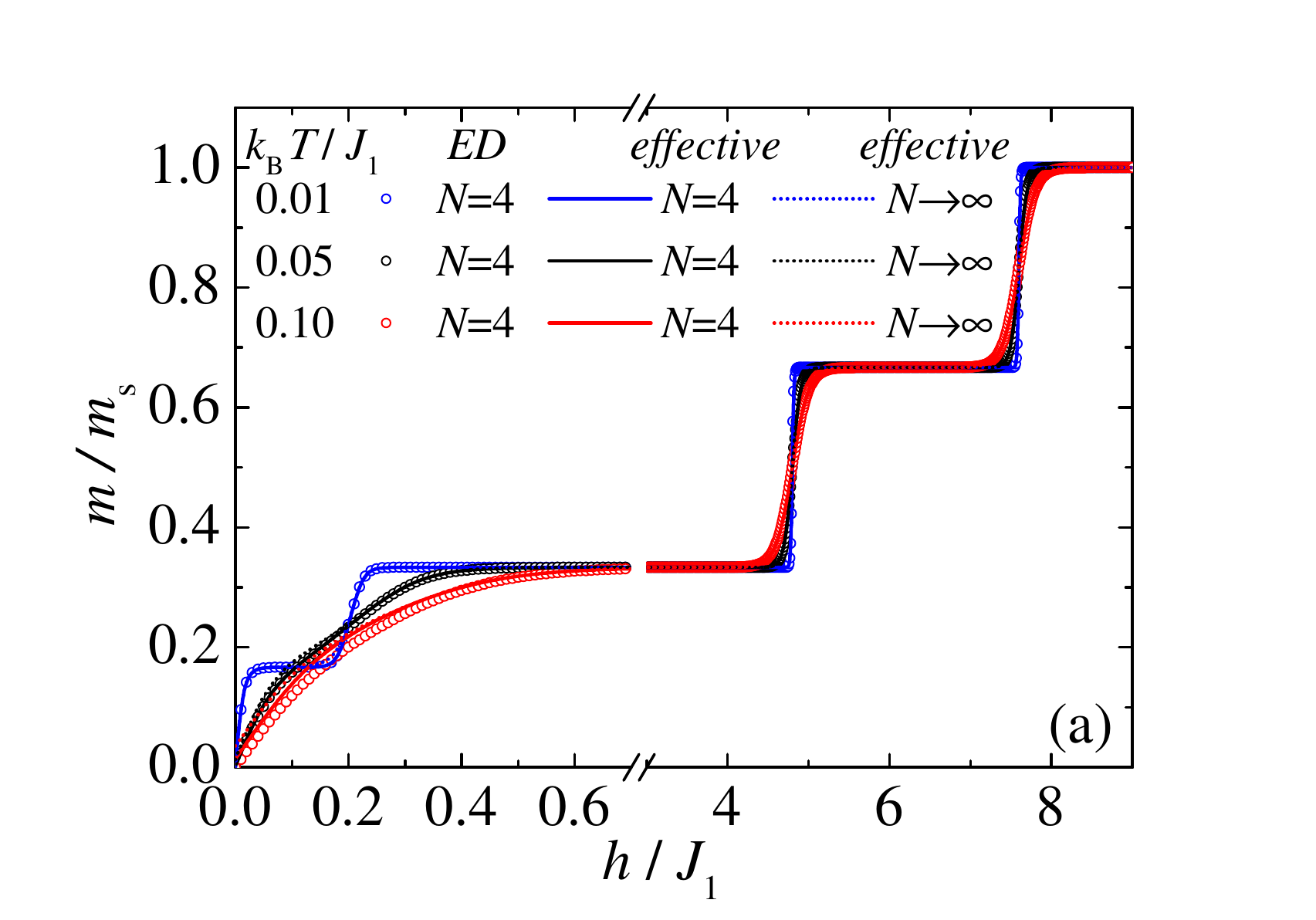} 
	\hspace*{-0.5cm}
	\includegraphics[width=0.5\textwidth]{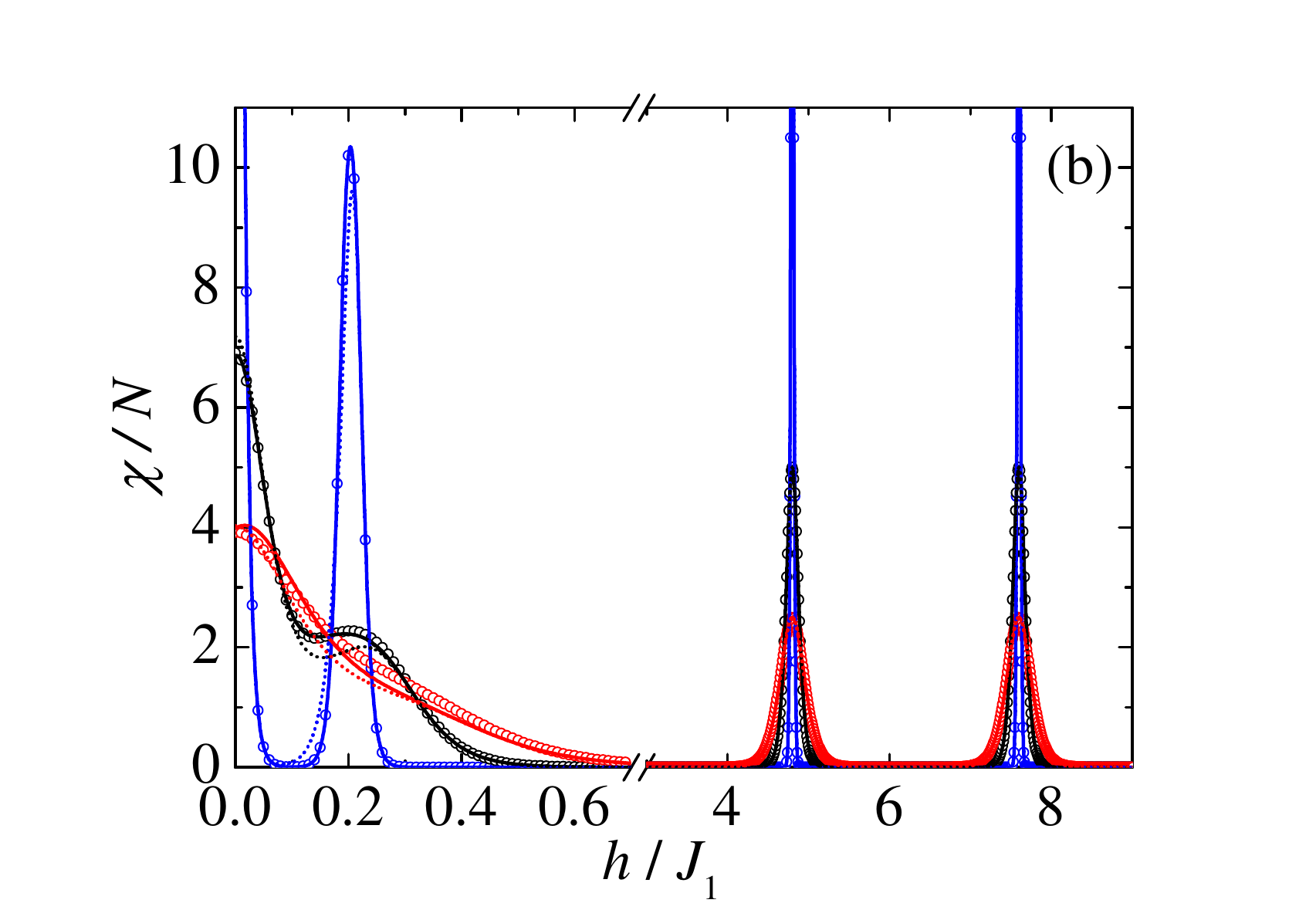}  
	\includegraphics[width=0.5\textwidth]{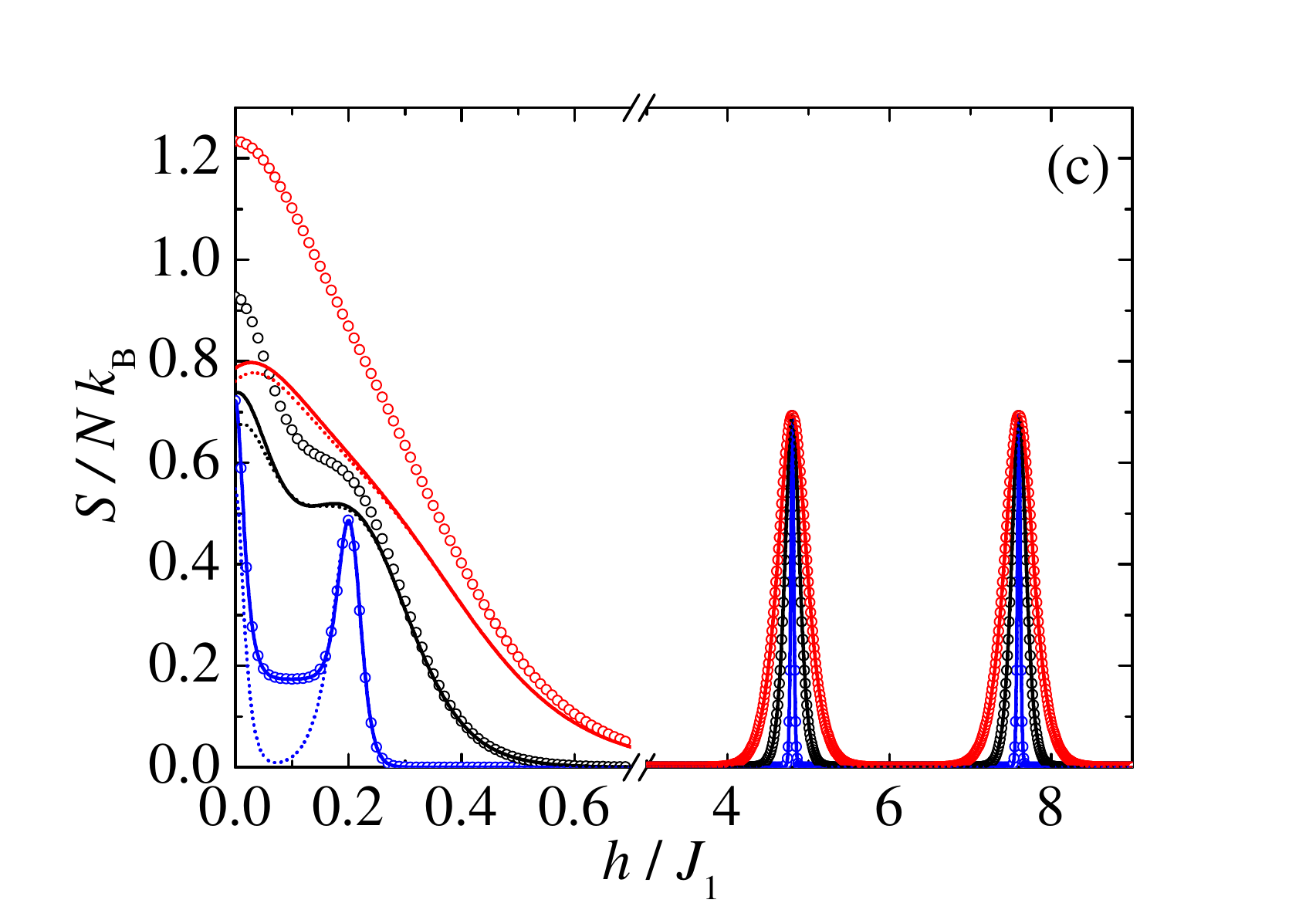} 
	\hspace*{-0.5cm}
	\includegraphics[width=0.5\textwidth]{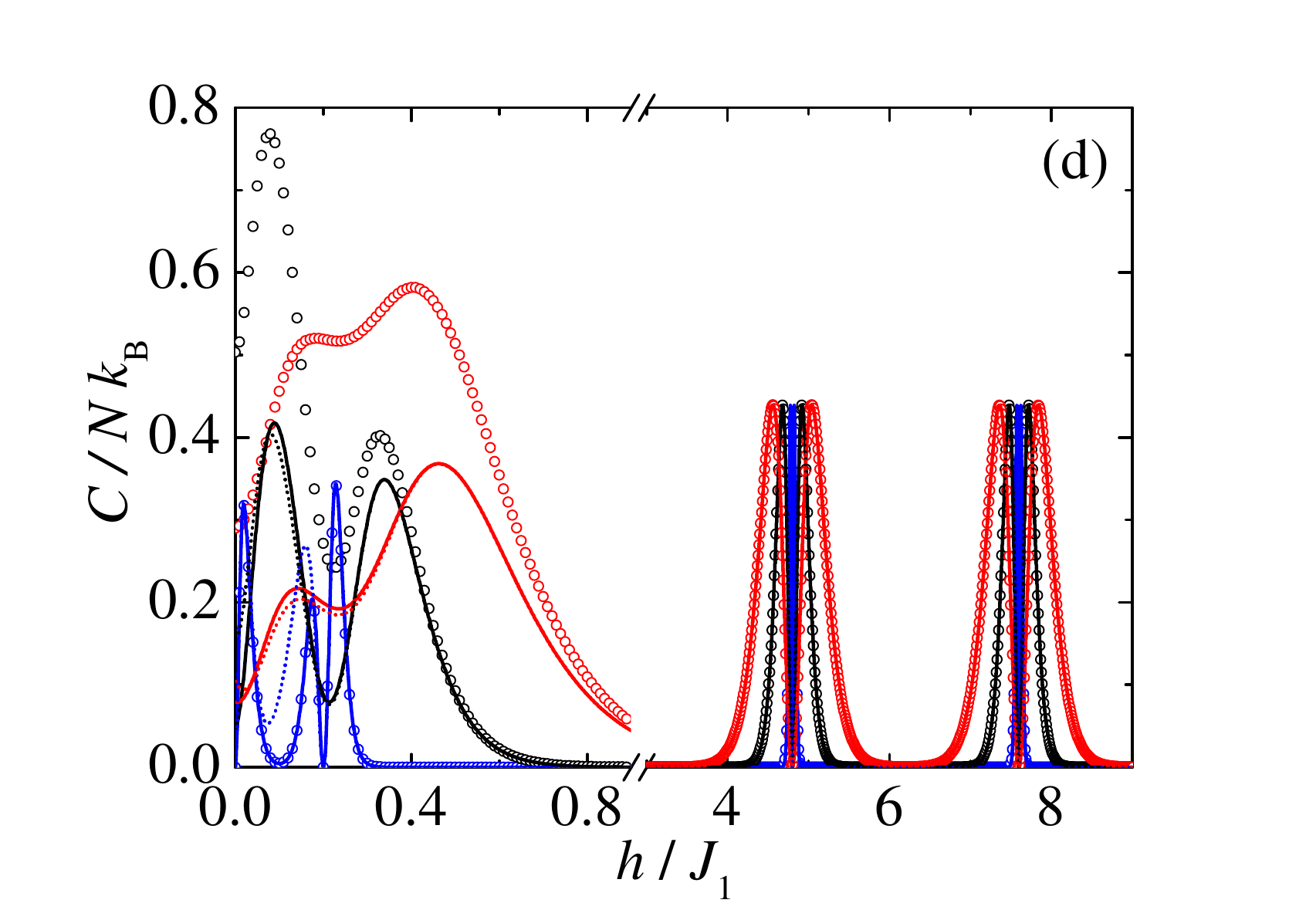} 
	\vspace{-0.5cm}  
	\caption{Isothermal magnetic-field dependencies of (a) the magnetization normalized to the saturation value $m/m_\text{s}$, (b) the magnetic susceptibility per unit cell $\chi/N$, (c) the entropy per unit cell $S/N k_{\rm B}$, and (d) the specific heat per unit cell $C/N k_{\rm B}$  for the spin-1 Heisenberg diamond chain at the fixed value of the interaction ratio $J_2/J_1 = 2.8$ and three selected temperatures: $k_{\rm B}T/J_1 = 0.01$ (blue), $k_{\rm B}T/J_1 = 0.05$ (black), and $k_{\rm B}T/J_1 = 0.1$ (red). Symbols denote ED data for a diamond chain with $N=4$ unit cells, while solid and dotted lines represents results of the effective lattice-gas model for $N=4$ and $N \to \infty$ unit cells, respectively. The legend shown in the panel (a) applies also for other panels.}
\label{fig:Mz_S_ED_LM_J28}
\end{figure*}

In Fig. \ref{fig:Mz_S_ED_LM_J28}, we compare analytical predictions from the effective monomer-dimer lattice-gas model (solid lines) with numerical ED data (symbols) for a finite spin-1 Heisenberg diamond chain of 12 spins (i.e. $N=4$ unit cells). As shown in Figs. \ref{fig:Mz_S_ED_LM_J28}(a)-(d), the magnetization, magnetic susceptibility, entropy, and specific heat obtained from the effective lattice-gas model perfectly agree with ED results at low enough temperatures $k_{\rm B} T/J_1 \lesssim 0.05$. This excellent quantitative agreement confirms the validity of the effective lattice-gas description in the parameter space highlighted in Fig. \ref{fig:GSPD}(a), from which the representative case with the interaction ratio $J_2/J_1 = 2.8$ was chosen for illustration. As the temperature further increases, however, small but systematic deviations emerge particularly in thermodynamic quantities such the entropy and specific heat in a weak-field regime $h/J_1 \lesssim 0.4$, whereas the discrepancies in magnetic quantities like the magnetization and magnetic susceptibility still remain minor. These results demonstrate that the effective monomer-dimer lattice-gas model provides a highly accurate description of low-temperature magnetic and thermodynamic properties of the spin-1 Heisenberg diamond chain, while still capturing most prominent features at moderate temperatures and weaker fields at least qualitatively. At higher temperatures, the deviations between the effective description and ED grow significantly, because the effective model fails to capture collective (non-fragmented) quantum states of the spin-1 Heisenberg diamond chain contributing to high-energy excitations. Having firmly established that the effective monomer-dimer lattice gas provides at low temperatures a reliable description of the finite-size spin-1 Heisenberg diamond chain, it is reasonable to conjecture that the effective description remain valid also in the thermodynamic limit $N \to \infty$ that is beyond the reach of any unbiased numerical method. The comparison between results derived within the effective lattice-gas description for $N=4$ and $N \to \infty $ indicates that finite-size effects are negligible at higher magnetic fields and become relevant only in the low-field region $h/J_1 \lesssim 0.4$. This behavior reflects the fact that the bound one- and two-magnon states encoded by two types of hard monomers dominate the low-energy spectrum at high and moderate magnetic fields, whereas triplet states of a diamond plaquette effectively encoded by hard dimers, which are predominantly responsible for finite-size effects, dominate the low-energy spectrum in the low-field regime. 

\begin{figure}  
	\centering
	\includegraphics[width=0.50\textwidth]{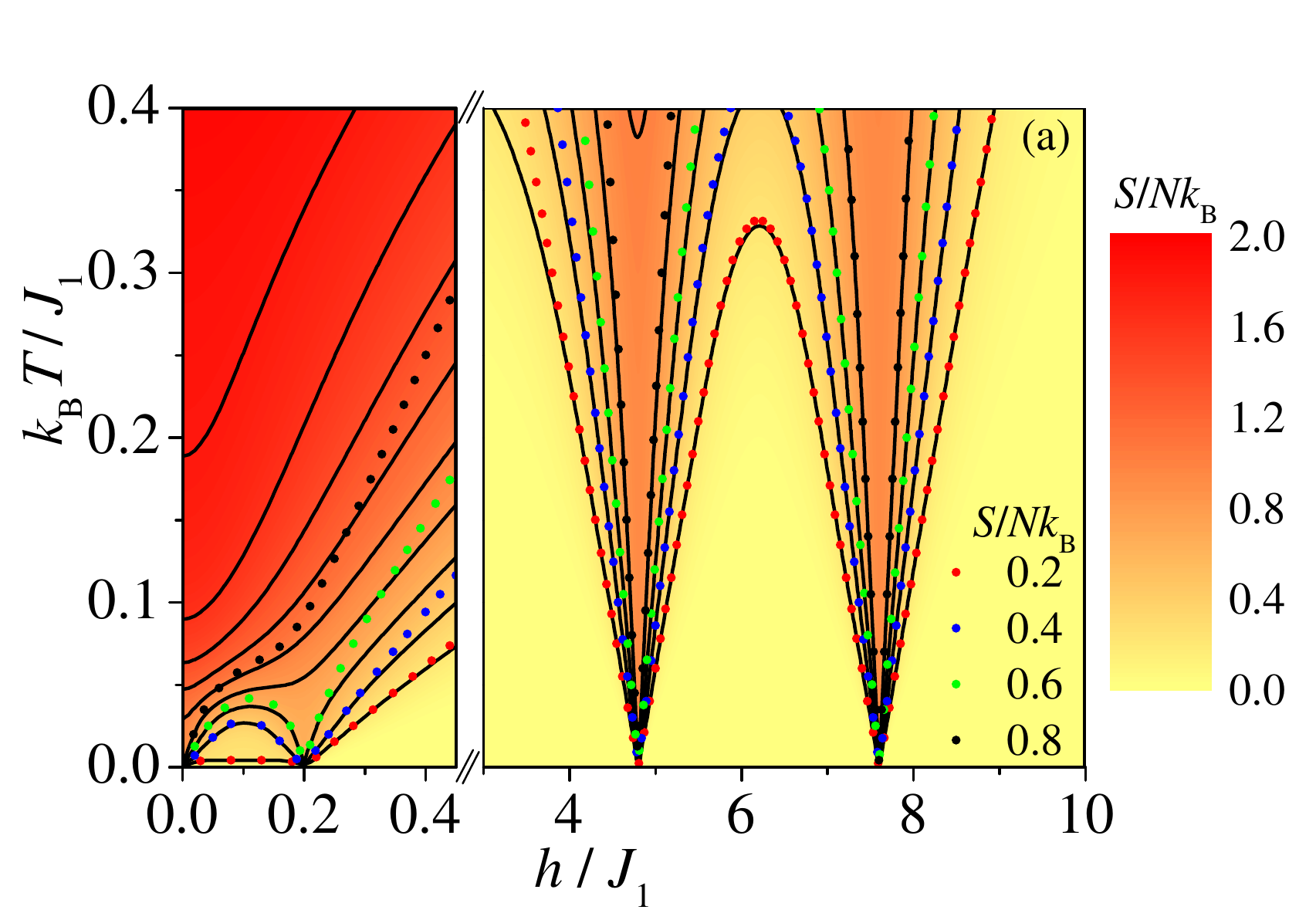}
	\includegraphics[width=0.50\textwidth]{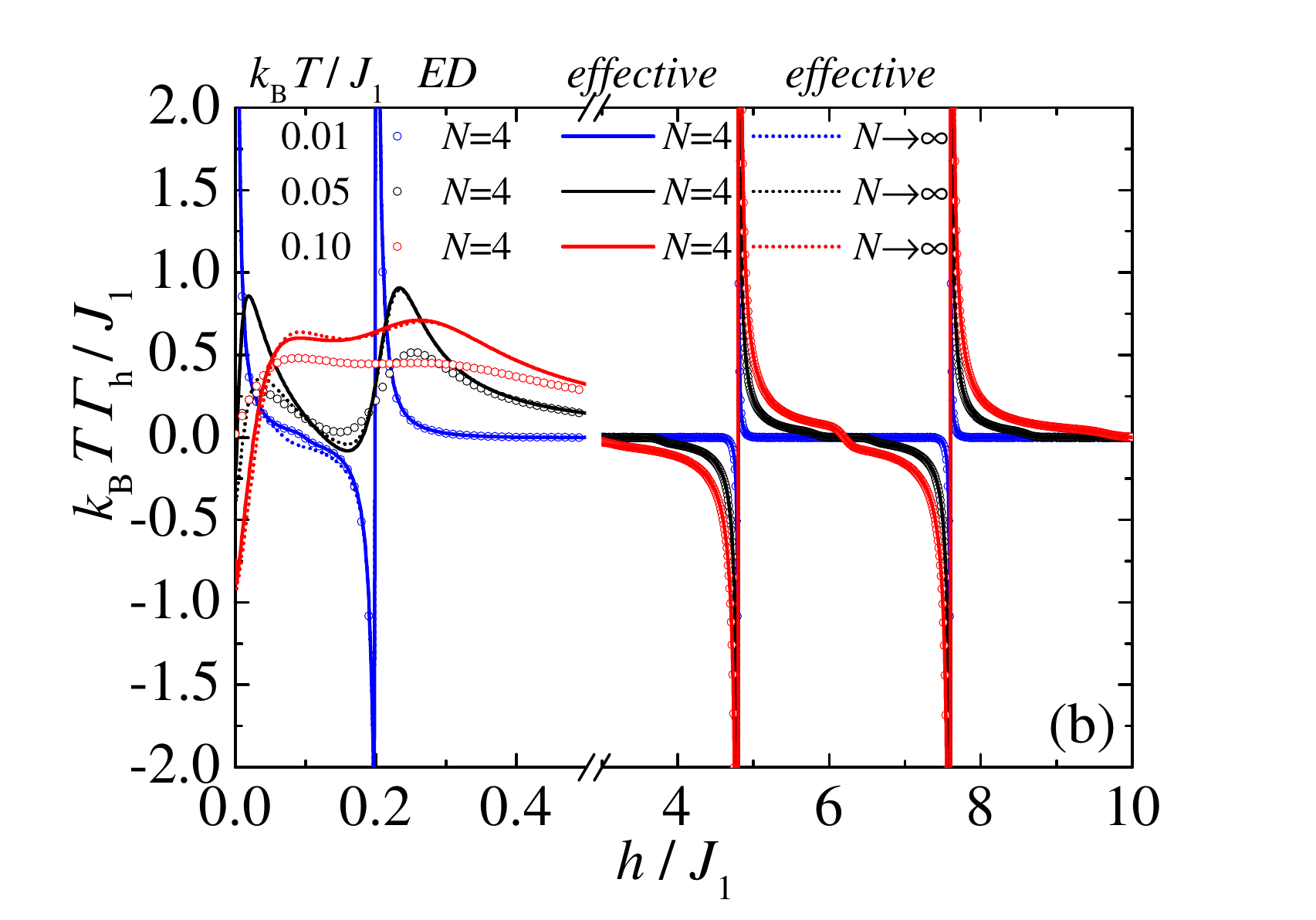}
	\vspace{-0.25cm}  
	\caption{(a) Density plot of the magnetic entropy of the spin-1 Heisenberg diamond chain for the interaction ratio $J_2/J_1 = 2.8$ and $N=4$ unit cells in the field-temperature plane calculated using the ED method. The ED data for the isentropic contour lines shown by solid lines are confronted with the results of effective lattice-gas description depicted by symbols for four entropic values illustrated in the legend; (b) The product of the magnetic Gr{\"u}neisen parameter with temperature as a function of the magnetic field for three temperatures $k_{\rm B} T/J_1 = 0.01$ (blue), 0.05 (black), and 0.1 (red). Symbols correspond to ED data for a spin-1 Heisenberg diamond chain with $N=4$ unit cells, while solid and dotted lines denote results obtained from the effective lattice-gas model with $N=4$ and $N\rightarrow \infty$ unit cells, respectively.}
	\label{fig:mce}
\end{figure}

Frustrated quantum magnets are well known for exhibiting an enhanced MCE during the adiabatic demagnetization, which makes them promising candidates for cost-effective magnetic refrigeration \cite{zhit03,gsch05}. One-dimensional frustrated quantum spin chains have likewise emerged as promising candidate materials for low-temperature magnetic refrigeration \cite{zhit04}. Let us therefore focus on the cooling capabilities of the frustrated spin-1 Heisenberg diamond chain. As a first step, we analyze temperature variations of the frustrated spin-1 Heisenberg diamond chain at a fixed interaction ratio $J_2/J_1=2.8$ under the adiabatic reduction of the applied magnetic field. Fig. \ref{fig:mce}(a) presents a density plot of the magnetic entropy $S/N k_{\rm B}$ in the magnetic field versus temperature plane together with several isentropic contour lines. These contour lines enable to trace temperature changes induced by sweeping the external magnetic field under the adiabatic condition. The density plot was obtained from ED of the spin-1 Heisenberg diamond chain with $N=4$ unit cells, whereby symbols mark four lowest-entropy contour lines calculated from the effective lattice-gas model. It is evident from Fig. \ref{fig:mce}(a) that a pronounced MCE characterized by abrupt cooling and heating achieved upon lowering the magnetic field can be detected in the high-field region in vicinity of the transition fields $h/J_1 = 7.6$ and $h/J_1 = 4.8$, which correspond to the field-driven transitions between FM-BMC and BMC-MD phases, respectively. A similarly strong MCE appears in the low-field region around zero field and near the transition field $h/J_1 = 0.2$ associated with the field-induced transition between MD and TD phases. The rapid magnetocaloric cooling near zero field is particularly appealing for potential technological applications, since only a weak driving field is required for substantial temperature drop due to the enhanced MCE.

To bring deeper insight into the adiabatic cooling rate, we next examine the typical behavior of the magnetic Gr{\"u}neisen parameter defined by the formula:
\begin{equation}
\label{Eq:Gamma}
\Gamma_h = \frac{1}{T}\left(\frac{\partial T}{\partial h}\right)_S.
\end{equation}
The product of the magnetic Gr{\"u}neisen parameter either with temperature ($T\Gamma_h$) or magnetic field ($h \Gamma_T$) serves as another key characteristics of MCE. Notably, the product 
$T\Gamma_h$ depicted in Fig. \ref{fig:mce}(b) as a function of the magnetic field directly determines the cooling rate during adiabatic demagnetization: a sign change of $T\Gamma_h$ from positive to negative values indicates the crossover from cooling to heating regime upon the adiabatic removal of the external magnetic field. The ED data for the spin-1 Heisenberg diamond chain with $N=4$ unit cells are shown by symbols, while solid and dotted lines were extracted from the effective monomer-dimer lattice-gas model with $N=4$ and $N\rightarrow \infty$ unit cells, respectively. A direct comparison  reveals that the effective model faithfully reproduces all essential features at low enough temperatures $k_{\rm B}T/J_1 \lesssim 0.05$. At a higher temperature $k_{\rm B}T/J_1 = 0.1$, the effective model still captures the overall trends in a high-field region $h/J_1 \gtrsim 0.5$ though more noticeable deviations appear in the low-field range $h/J_1 \lesssim 0.5$. The field-driven phase transitions emergent in the spin-1 Heisenberg diamond chain at zero temperature are reflected also by pronounced peaks and sign reversals of the adiabatic cooling rate at low but nonzero temperatures. While the behavior of $T\Gamma_h$ product is largely unaffected by finite-size effects near the transitions fields  $h/J_1 = 7.6$ and $h/J_1 = 4.8$ emerging in a high-field regime, the pronounced finite-size effects can be repeatedly observed in the low-field region especially around zero field and the lowest transition field $h/J_1 = 0.2$. 

\section{Quantum Stirling heat engine}
\label{QSHE}

\begin{figure}
	\centering
	\includegraphics[width=0.5\textwidth]{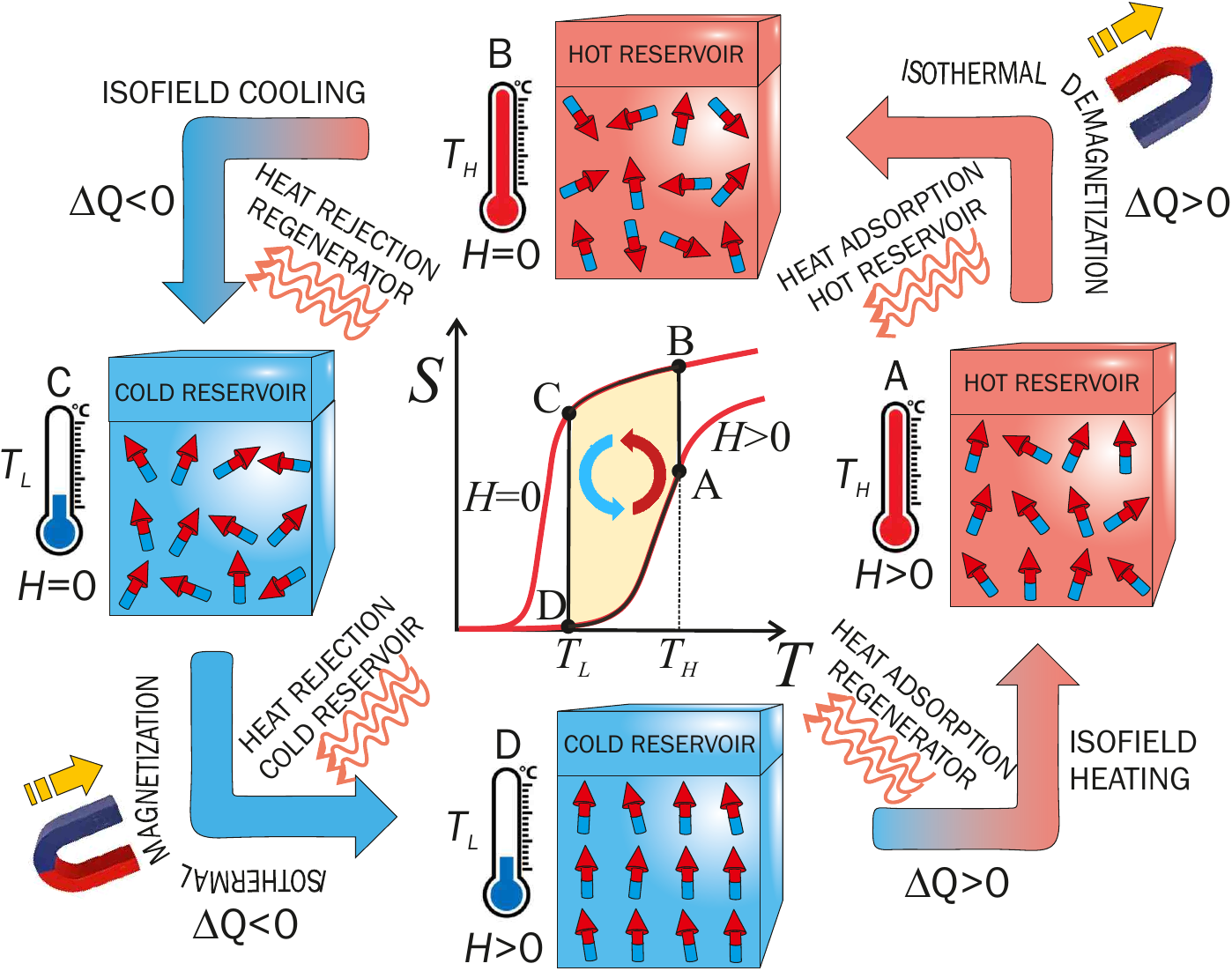}
	\vspace{-0.25 cm}
	\caption{A schematic diagram of the quantum Stirling cycle. The central part shows temperature variations of the entropy at two distinct magnetic fields including four stages and processes of the cycle. The process A $\rightarrow$ B corresponds to the isothermal demagnetization when the magnetic field reduces from $h_H>0$ to $h_L=0$, the stroke B $\rightarrow$ C corresponds to the isofield cooling when temperature decreases from $T_H$ to $T_L$, the process C $\rightarrow$ D corresponds to the isothermal magnetization invoked by increasing of the magnetic field from $h_L=0$ to $h_H>0$, while the last stroke D $\rightarrow$ A is the isofield heating associated with temperature increase from $T_L$ to $T_H$.}
	\label{fig:StirlingC}
\end{figure}

The efficient conversion of thermal energy into mechanical work by a heat engine is an ongoing technological challenge with the Carnot limit setting the ultimate theoretical bound on efficiency. The quantum Stirling cycle represents an intriguing quantum analogue of its classical counterpart, which consists of two isothermal and two isofield strokes as illustrated in Fig. \ref{fig:StirlingC}. The  magnetic Stirling cycle begins with the isothermal demagnetization A $\rightarrow$ B, during which the working substance is in thermal equilibrium with a hot reservoir at temperature $T_H$ while the external magnetic field is quasi-statically reduced from $h_H$ to $h_L$. As a result, the system performs work and absorbs heat $Q_{AB} = T_H [S_B (T_H, h_L) - S_A (T_H, h_H)]>0$. In the second step B $\rightarrow$ C, the system undergoes isofield cooling when its temperature decreases from $T_H$ to $T_L$ due to the heat release $Q_\text{BC} = U_C (T_L, h_L) - U_B (T_H, h_L) <0$ to the regenerator acting as an internal heat storage unit of the engine while no work is being done during this process. During the isothermal magnetization serving as the third step C $\rightarrow$ D, the working substance is in thermal equilibrium with a cold reservoir at temperature $T_L$ while being gradually magnetized by increasing the external magnetic field back from $h_L$ to $h_H$. Note that the external work is required to isothermally magnetize the working substance, while the heat $Q_{CD} = T_L [S_D(T_L, h_H) - S_C(T_L, h_L)]<0$ is released to the cold reservoir. The last step of the cycle D $\rightarrow$ A involves the isofield heating of the working substance when its temperature increases from $T_L$ to $T_H$ at constant magnetic field $h_H$. This happens due to the absorption of the heat $Q_\text{DA} = U_A (T_H, h_H)- U_D (T_L, h_H) >0$ from the regenerator without performing external work. The net work done over the entire Stirling cycle is then simply obtained from the first law of thermodynamics as the sum of heat exchanges during all four stages: 
\begin{eqnarray}
W &=& Q_{AB} + Q_{BC} + Q_{CD} + Q_{DA} \nonumber \\ 
  &=& T_H [S_B (T_H, h_L) - S_A (T_H, h_H)] \nonumber \\
	&+& U_C (T_L, h_L) - U_B (T_H, h_L) \nonumber \\
	&+& T_L [S_D(T_L, h_H) - S_C(T_L, h_L)] \nonumber \\
	&+& U_A (T_H, h_H) - U_D (T_L, h_H).
\end{eqnarray}  
If the system operates as a quantum heat engine, the efficiency $\eta$ of converting heat into mechanical work can be then defined as the ratio of the total work done to the heat absorbed from the hot reservoir:
\begin{equation}
\eta = \frac{W}{Q_{AB}} = \frac{W}{T_H [S_B (T_H, h_L) - S_A (T_H, h_H)]}.
\end{equation}  

\begin{center}
	\begin{figure}
		\includegraphics[width=0.5\textwidth]{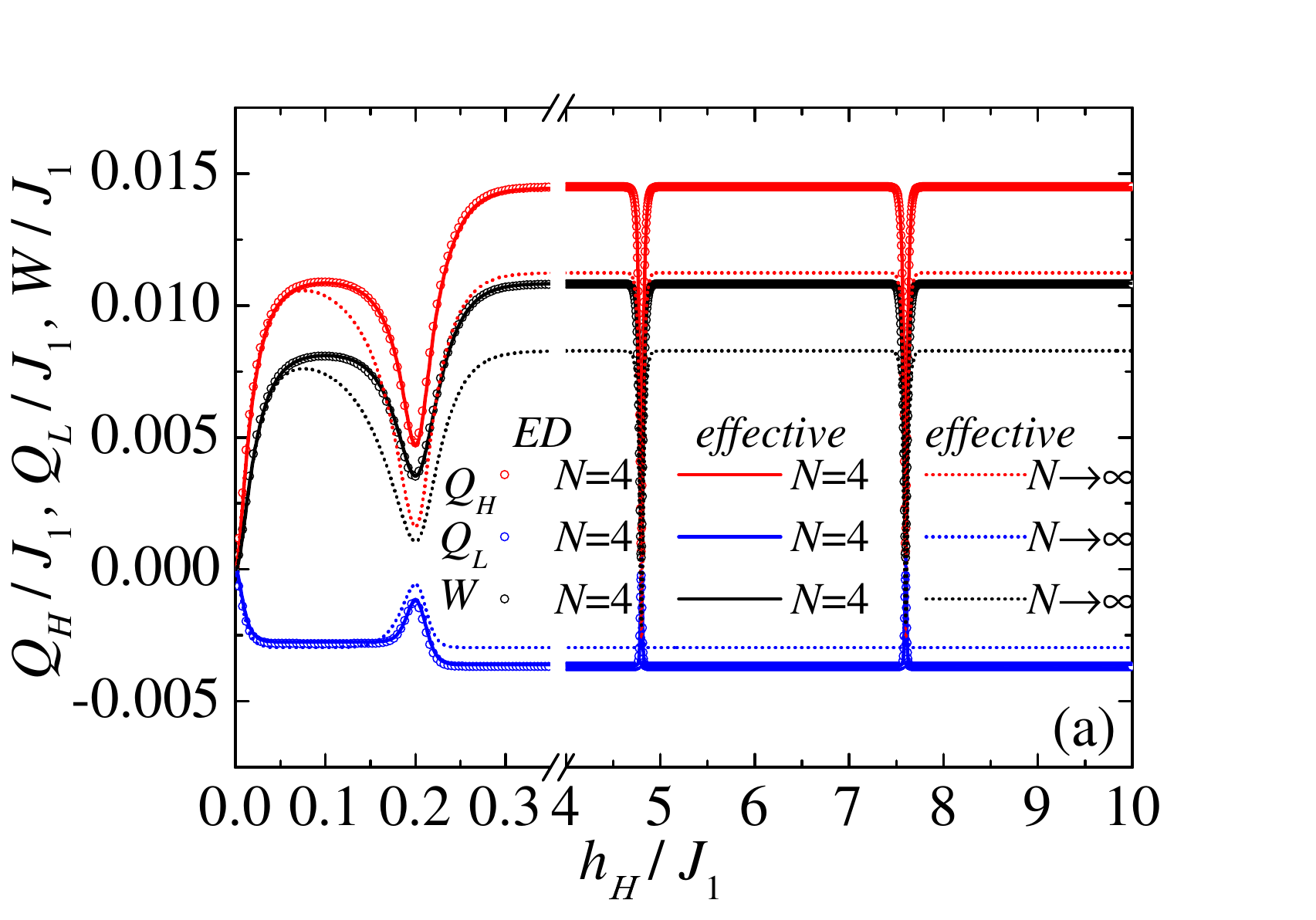} 
		\includegraphics[width=0.5\textwidth]{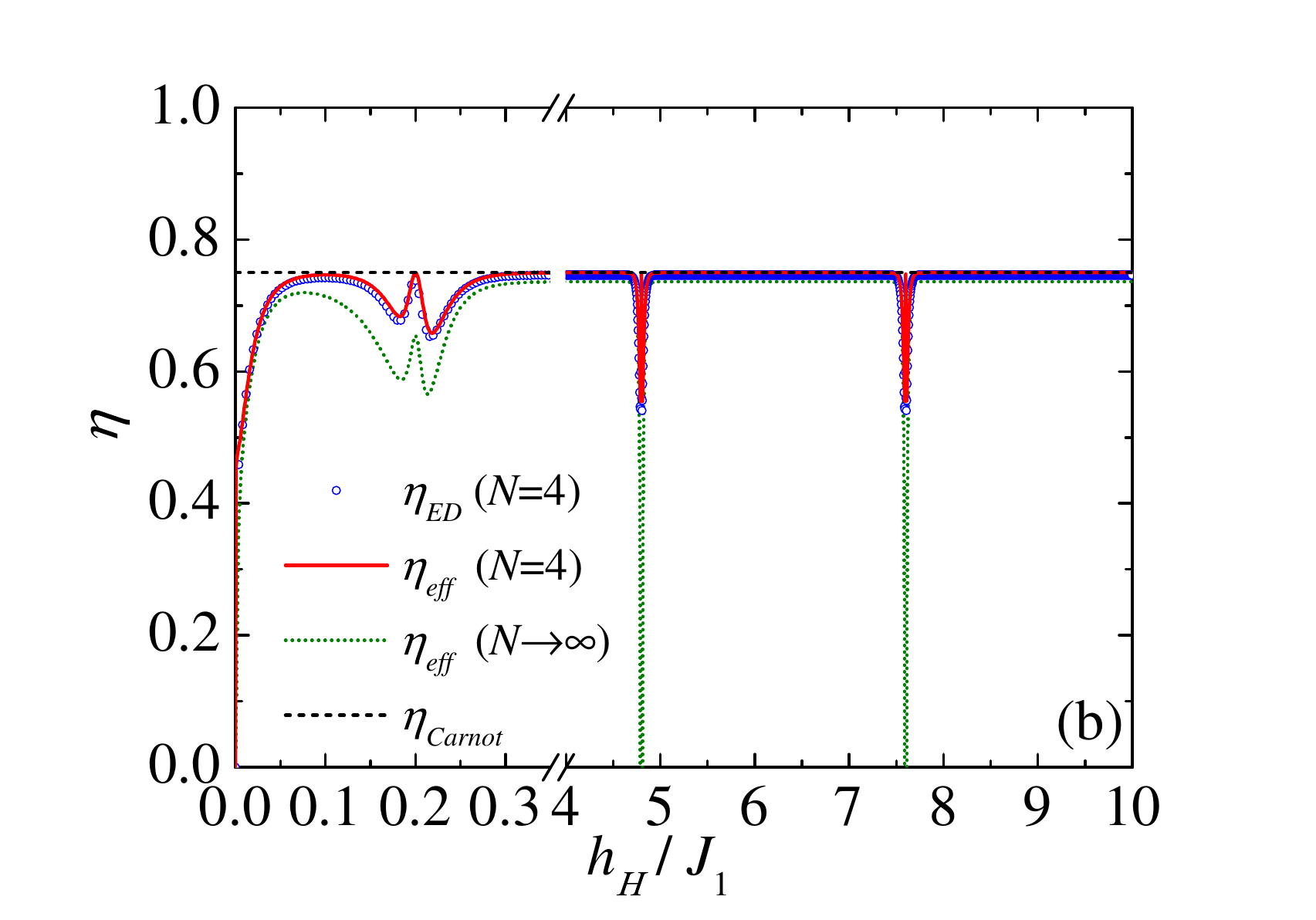}
		\vspace{-0.25 cm}
		\caption{(a) The net output work $W$ and the heat exchanges $Q_L$ and $Q_H$ as obtained from the spin-1 Heisenberg diamond chain with the ratio $J_2/J_1 = 2.8$ as the working medium of the quantum Stirling cycle, which is composed from two isothermal processes held at temperatures $k_{\rm B}T_L/J_1 = 0.005$ and $k_{\rm B}T_H/J_1 = 0.02$ and two isofield processes kept at zero field $h_L/J_1 = 0.0$ and variable upper field $h_H/J_1$. Symbols denote ED data for a diamond chain with $N = 4$ unit cells, while solid and dotted lines represents results of the effective lattice-gas model for $N = 4$ and $N  \to \infty$ unit cells, respectively; (b) The efficiency of the aforementioned quantum Stirling cycle and its comparison with the upper Carnot limit.} 
		\label{fig:QHE}
	\end{figure}
\end{center}

The magnetic Stirling cycle offers efficient energy conversion by a careful manipulation of magnetic-field strengths as well as temperatures of hot and cold thermal reservoirs, which may allow tuning an efficiency of the quantum heat engine close to an upper bound determined by the Carnot limit. Let us therefore consider the frustrated spin-1 Heisenberg diamond chain with the coupling constant $J_2/J_1 = 2.8$ as the working medium of the quantum Stirling heat engine by assuming zero field $h_L=0$ during the isofield cooling and variable upper field $h_H>0$ during the isofield heating. Three key thermodynamic characteristics of the magnetic Stirling cycle -- the absorbed heat $Q_H$ from a thermal reservoir at a higher temperature $k_{\rm B} T_H/J_1 = 0.02$, the heat released $Q_L$ to a thermal reservoir at a lower temperature $k_{\rm B} T_L/J_1 = 0.005$, and the net work output $W$ of the whole cycle are presented in Fig. \ref{fig:QHE}(a) as a function of the upper magnetic field. To assess finite-size effects, ED data for the spin-1 Heisenberg diamond chain with $N=4$ unit cells (symbols) are confronted with results from the effective monomer-dimer lattice-gas model with $N=4$ (solid lines) and $N \to \infty$ (dotted lines) unit cells, respectively. The near-perfect agreement between ED data and effective results for $N=4$ unit cells demonstrates that operating the magnetic Stirling cycle within the temperature range $0.005 \leq k_{\rm B} T/J_1 \leq 0.02$ ensures the acceptable accuracy and consistency of both sets of results within the specified temperature range.

It is clear from Fig. \ref{fig:QHE}(a) that the frustrated spin-1 Heisenberg diamond chain acts over a broad range of upper magnetic fields as an efficient working medium for the magnetic Stirling heat engine as $W>0$, $Q_H>0$, and $Q_L<0$. The amount of heat exchanged with both thermal reservoirs as well as the net work output indeed remain nearly constant across most of the magnetic-field range except those in vicinity of three transition fields $h_H/J_1 = 0.2$, $4.8$, and $7.6$, at which three field-driven phase transitions TD $\leftrightarrow$ MD, MD $\leftrightarrow$ BMC, and BMC $\leftrightarrow$ FM occur. It actually turns out that the net work, absorbed and released heat suddenly drop near all three field-driven phase transition. Another intriguing feature is that the amount of exchanged heat and extractable net work are slightly smaller when the upper magnetic field drives the diamond chain with $N=4$ unit cells into the cluster-based Haldane TD phase than the fully fragmented MD phase or the BMC phase. In the thermodynamic limit $N \to \infty$, the amount of exchanged heat and net output work approach nearly identical values regardless of whether the system is driven to the TD, MD, or BMC phase indicating that the engine’s performance becomes insensitive to the specific phase hosting the working medium once finite-size effects are eliminated. 

Finally, we present in Fig. \ref{fig:QHE}(b) the efficiency $\eta$ of the magnetiic Stirling heat engine with the frustrated spin-1 Heisenberg diamond chain serving as the working medium. It is demonstrated that the efficiency of the quantum engine approaches for most magnetic fields the ideal Carnot limit $\eta_{Carnot} = 1 - T_\text{L}/T_\text{H} = 0.75$, whereas it only significantly drops  when the magnetic field of the isofield heating process lies near the three transition fields $h_H/J_1 = 0.2$, $4.8$, and $7.6$ associated with TD-MD, MD-BMC, and BMC-FM transitions, respectively. The efficiency actually displays in a rather narrow field range centered around each transition field a characteristic double-well profile with a distinct maximum at a given transition field. The optimal operating condition for maximizing efficiency of the heat engine are thus found when the upper magnetic field of the isofield heating process is tuned to the middle of the field range of TD, MD, and BMC phases, i.e. $h_H/J_1 = 0.1$, $2.5$ and $6.2$ for $J_2/J_1 = 2.8$, when the efficiency approaches near-Carnot performance. 

\section{Conclusions}
\label{CC}

In this work, we have conducted a comprehensive study of the magnetic and thermodynamic properties of the spin-1 Heisenberg diamond chain in an external magnetic field by combining advanced analytical and numerical approaches. The system hosts a broad spectrum of exotic quantum phases including the Lieb-Mattis ferrimagnetic phase, uniform and cluster-based Haldane states, fully fragmented monomer-dimer phase, quantum spin liquids, and bound-magnon crystal alongside the trivial fully polarized phase. Although these unconventional quantum phases are not realized in the nickel-based coordination polymer [Ni$_3$(OH)$_2$(C$_4$H$_2$O$_4$)(H$_2$O)$_4$]·2H$_2$O due to its placement in the unfrustrated parameter regime supporting the Lieb-Mattis ferrimagnetic ground state, our findings establish the spin-1 diamond chain as a highly versatile platform for exploring field-induced quantum phase transitions, enhanced magnetocaloric effect, and working media for future quantum heat-engine technologies.

In the highly frustrated regime, we accurately captured the multistep magnetization curves and key thermodynamic properties such as entropy, specific heat, adiabatic cooling rate and work extraction in a quantum heat engine by mapping the spin-1 Heisenberg diamond chain onto an effective monomer-dimer lattice-gas model. The analytical predictions derived from this effective model show excellent quantitative agreement with ED data for finite spin-1 Heisenberg diamond chains, thereby confirming the validity and accuracy of the lattice-gas description in capturing low-temperature magnetic and thermodynamic behavior at low temperatures across the full range of magnetic fields. 

Our results reveal that experimental realizations the frustrated spin-1 Heisenberg diamond chain would be highly desirable as they provide efficient and tunable platform for both magnetocaloric refrigeration and quantum heat-engine technologies. In particular, this frustrated spin chain exhibits an enhanced MCE near field-driven phase transitions of the unconventional quantum phases such as the fragmented monomer-dimer, bound magnon crystal, and cluster-based Haldane phases, and notably high efficiency of the associated quantum heat engine when the applied magnetic field is tuned away from the respective transition fields. These findings pave the way for future research and development of quantum technologies that exploit the frustrated spin-1 diamond chain as an active working medium in adiabatic magnetic refrigeration and quantum heat Stirling engine.  

\section*{Acknowledgments}
Funded by the EU NextGenerationEU through the Recovery and Resilience Plan for Slovakia under the project Nos. 09I03-03-V04-00403 and 09I03-03-V05-00008 (VVGS-2023-2888), and The Slovak Research and Development Agency under the grant No. APVV-24-0091. 

\section*{Data Availability Statement}
All data generated or analyzed during this study are included in this published article. Additional numerical data supporting the findings of this work are available from the corresponding author upon reasonable request.

\section*{Author Contributions}
J.S., A.Z., and K.K. conceived the research project. A.Z., H.A.Z., and J.S. performed analytical calculations and developed the effective lattice-gas model. K.K. carried out DMRG simulations and constructed the ground-state phase diagram. J.S. performed  QMC simulations. A.Z. and J.S. performed the exact diagonalization and prepared the figures. All authors discussed the results, contributed to the interpretation of the data. A.Z. and H.A.Z. contributed to the first draft of the manuscript, J.S. finalized the manuscript. All authors reviewed the manuscript.

\end{document}